\documentclass[twocolumn]{aastex631}
\usepackage{multirow}
\usepackage{amssymb}
\usepackage{amsmath}
\usepackage{tablefootnote}
\usepackage{xcolor}

\shorttitle{HD~134606}
\shortauthors{Li et al.}

\newcommand{\shk}{$S_{\mathrm{HK}}$~}
\newcommand{\logrphk}{log\,$R^{\prime}_{HK}$}

\begin{document}

\title{Revised Architecture and Two New Super-Earths in the HD~134606 Planetary System}

\author[0000-0002-4860-7667]{Zhexing Li}
\affiliation{Department of Earth and Planetary Sciences, University of California, Riverside, CA 92521, USA}
\email{zli245@ucr.edu}

\author[0000-0002-7084-0529]{Stephen R. Kane}
\affiliation{Department of Earth and Planetary Sciences, University of California, Riverside, CA 92521, USA}

\author[0000-0003-2630-8073]{Timothy D.~Brandt}
\affiliation{Department of Physics, University of California, Santa Barbara, CA 93106, USA}

\author[0000-0002-3551-279X]{Tara Fetherolf}
\affiliation{Department of Physics, California State University, San Marcos, CA 92096, USA}
\affiliation{Department of Earth and Planetary Sciences, University of California, Riverside, CA 92521, USA}

\author[0000-0003-0149-9678]{Paul Robertson}
\affiliation{Department of Physics \& Astronomy, University of California, Irvine, CA 92697, USA}

\author[0000-0001-5290-2952]{Jinglin~Zhao}
\affiliation{DTU Space, National Space Institute, Technical University of Denmark, Elektrovej 328, DK-2800 Kgs. Lyngby, Denmark}
\affiliation{Department of Astronomy \& Astrophysics, The Pennsylvania State University, 525 Davey Lab, University Park, PA 16802, USA}

\author[0000-0002-4297-5506]{Paul A.\ Dalba}
\altaffiliation{Heising-Simons 51 Pegasi b Postdoctoral Fellow}
\affiliation{Department of Astronomy and Astrophysics, University of California, Santa Cruz, CA 95064, USA}

\author[0000-0001-9957-9304]{Robert A. Wittenmyer}
\affiliation{University of Southern Queensland, Centre for Astrophysics, West Street, Toowoomba, QLD 4350 Australia}

\author[0000-0003-1305-3761]{R. Paul Butler}
\affiliation{Earth and Planets Laboratory, Carnegie Institution for Science, 5241 Broad Branch Road NW, Washington DC 20015-1305, USA}

\author[0000-0002-2100-3257]{Matías R. Díaz}
\affil{Las Campanas Observatory, Carnegie Institution of Washington, Colina El Pino, Casilla 601 La Serena, Chile}

\author[0000-0002-2532-2853]{Steve~B.~Howell}
\affil{NASA Ames Research Center, Moffett Field, CA 94035, USA}

\author[0000-0002-5726-7000]{Jeremy Bailey}
\affil{Exoplanetary Science at UNSW, School of Physics, UNSW Sydney, NSW, 2052, Australia}

\author[0000-0003-0035-8769]{Brad Carter}
\affil{University of Southern Queensland, Centre for Astrophysics, West Street, Toowoomba, QLD 4350 Australia}

\author[0000-0001-9800-6248]{Elise~Furlan}
\affiliation{NASA Exoplanet Science Institute, Caltech/IPAC, Mail Code 100-22, 1200 E. California Blvd., Pasadena, CA 91125, USA}

\author[0000-0003-2519-6161]{Crystal~L.~Gnilka}
\affil{NASA Ames Research Center, Moffett Field, CA 94035, USA}

\author[0000-0003-0433-3665]{Hugh R.A. Jones}
\affil{Centre for Astrophysics Research, University of Hertfordshire, College Lane, Hatfield AL10 9AB, UK}

\author[0000-0003-2839-8527]{Simon O'Toole}
\affil{Australian Astronomical Optics, Macquarie University, 105 Delhi Rd, North Ryde NSW 2113, Australia}

\author[0000-0002-7595-0970]{Chris Tinney}
\affil{Exoplanetary Science at UNSW, School of Physics, UNSW Sydney, NSW, 2052, Australia}


\begin{abstract}

Multi-planet systems exhibit a diversity of architectures that diverge from the solar system and contribute to the topic of exoplanet demographics. Radial velocity (RV) surveys form a crucial component of exoplanet surveys, as their long observational baselines allow searches for more distant planetary orbits. This work provides a significantly revised architecture for the multi-planet system HD~134606 using both HARPS and UCLES RVs. We confirm the presence of previously reported planets b, c, and d with periods $12.0897^{+0.0019}_{-0.0018}$, $58.947^{+0.056}_{-0.054}$, and $958.7^{+6.3}_{-5.9}$ days, and masses $9.14^{+0.65}_{-0.63}$, $11.0\pm1$, and $44.5\pm2.9$ Earth masses respectively, with the planet d orbit significantly revised to over double that originally reported. We report two newly detected super-Earths, e and f, with periods $4.31943^{+0.00075}_{-0.00068}$ and $26.9^{+0.019}_{-0.017}$ days, and masses $2.31^{+0.36}_{-0.35}$ and $5.52^{+0.74}_{-0.73}$ Earth masses, respectively. In addition, we identify a linear trend in the RV time series, and the cause of this acceleration is deemed to be a newly detected sub-stellar companion at large separation. HD~134606 now displays four low mass planets in a compact region near the star, one gas giant further out in the Habitable Zone, an additional massive companion in the outer regime, and a low mass M dwarf stellar companion at large separation, making it an intriguing target for system formation/evolution studies. The location of planet d in the Habitable Zone proves to be an exciting candidate for future space-based direct imaging missions, whereas continued RV observations of this system are recommended for understanding the nature of the massive, long period companion.

\end{abstract}

\keywords{planetary systems -- techniques: photometric -- techniques: radial velocities -- stars: individual (HD~134606)}


\section{Introduction}
\label{sec:intro}

Exoplanet discoveries have revealed a vast range of planetary architectures, many of which differ significantly from that seen in the solar system \citep{ford2014,winn2015,horner2020b,kane2021d,mishra2023a,mishra2023b}. The expansion of exoplanet discovery space into the outer regions of planetary systems has been strongly enabled by the improving precision and increasing baseline of radial velocity (RV) surveys \citep{fischer2016}. Such surveys allowed increased RV sensitivities towards the larger semimajor axis regime and provided insights into the prevalence of Jupiter analogs
\citep{wittenmyer2011a,wittenmyer2020b,fulton2021,rosenthal2021}, indicating that, for orbits of 1--10~AU, $\sim$6\% of solar-type stars harbor giant planets \citep{wittenmyer2016c}. Further out, although companion orbits cannot be fully resolved due to incomplete RV sampling, their presence can usually be hinted in the form of a linear or quadratic acceleration on top of already recovered RV signals. 

At the other end of the spectrum, another benefit brought by some of these RV surveys is their capabilities of probing the low-mass short-period planetary regime if such surveys were conducted on a high cadence schedule with high precision spectrographs. Smaller planets induce lower RV semiamplitudes from the stellar reflex motion close to the noise floor. As such, these planets often require a lot more sampling over many orbital periods to gradually build up their statistical significance. With long running time and high precision, some of the RV surveys, or multiple ones combined, opened the door for discoveries of multiple planets with drastically different orbital properties within the same system. These multi-planet systems provide an essential contextual platform from which to evaluate formation scenarios within a statistical framework, and determine common versus rare outcomes of planetary architectures \citep{chambers1996,lissauer2011b,tremaine2012}. Furthermore, planetary systems with bright, nearby host stars are essential targets for direct imaging surveys that enable spectroscopic studies of planetary atmospheres to be undertaken \citep{kane2013c,clanton2016,kopparapu2018,nielsen2019c,stark2020,kane2022b,quanz2022b}. Thus, the discovery of nearby multi-planet systems, particularly those that harbor planets of different characteristics, are of enormous value in determining their contribution to exoplanet demographical studies and the potential for significant follow-up opportunities.

The HD~134606 system presents itself as one such exciting target. The system was reported by \citet[M11 hereafter]{mayor2011} to host three exoplanet candidates using RV data acquired from the High Accuracy Radial velocity Planet Searcher (HARPS) spectrograph \citep{pepe2000}. The planetary system was presented as part of a larger catalog of exoplanet candidates detected from HARPS observations. At the time, the three planets were reported to have orbital periods of 12, 59, and 459 days. Over 12 years since the original report, we dive back into the system equipped with a much extended HARPS observation baseline complemented by RVs from the University College London Echelle Spectrograph (UCLES) \citep{diego1990} to fully explore the system architecture and to check the validity of the originally reported candidates. Aliases and other false positives present in the analysis of RV data can often result in incorrect periodic signature extraction, and significantly impact the inferred system architecture \citep{dawson2010}. In addition, it is likely that further planets may be present in the system that remain undetected due to the precision of the RV data \citep{kane2007a,laliotis2023,newman2023}. In these cases, a planet injection/recovery methodology may be applied to determine the limits to which currently known planets represent the true system inventory \citep{howard2016}.

In this paper, we present a significantly revised architecture for the HD~134606 system, including a total of five planets and indications of a long-period additional companion. Section~\ref{sec:obs} provides descriptions of the various observations conducted and data sources used in our work, including RV, imaging, astrometry, and photometry. In Section~\ref{sec:analysis}, we present a full analysis of the system, including updated stellar parameters, a new RV model, and constraints on the additional companion within the system. Section~\ref{sec:activitysol} describes our study of stellar activity signals within the data, using both spectral activity indicators and photometry to explore the possibility of false positive signals within the RV data. Section~\ref{sec:discussion} discusses the various periodic signals in the data, aliases, dynamical stability, and the prospects for direct imaging. Our results are summarized in Section~\ref{sec:conclusions}, along with suggestions for future work.


\section{Observations and Data}
\label{sec:obs}

\subsection{Radial Velocities}
\label{sec:rvobs}
\subsubsection{HARPS}

The original discovery of the system by M11 utilized 113 RV observations by the HARPS spectrograph spanning a total of 2,548 days from July 2004 to July 2011. Since then, HARPS continued its monitoring of HD~134606 until May 2017. The additional observations significantly increased the number of RVs for the star to 219 and extended the baseline coverage to 4,677 days, or roughly 13 years. The extended coverage of the HARPS dataset includes a fibre upgrade for the instrument that resulted in a discontinuous jump in the RV time series around June 2015 \citep{locurto2015}. This full HARPS RV dataset was later re-analyzed and re-reduced as part of the new reduction for all the public HARPS spectra by \citet{trifonov2020} using the SpEctrum Radial Velocity AnaLyser (SERVAL) pipeline \citep{zechmeister2018}. The new reduction with the SERVAL pipeline corrected several systematics including nightly zero-point RVs and average intra-night drifts that slightly improves the precision of the HARPS RVs compared to those previously derived. The high observation cadence of HARPS and the improved precision thanks to the new reduction allowed for detection and characterization of possible short-period and low-semiamplitude exoplanet signals. The newly reduced HARPS RVs were published in the HARPS RV database \citep{trifonov2020} and we make use of those data in our work. 

\subsubsection{UCLES}

In addition to the HARPS data, we obtain RVs for HD~134606 as part of the Anglo-Australian Planet Search (AAPS) \citep{tinney2001} taken by the UCLES spectrograph mounted at the Anglo-Australian Telescope. AAPS is one of the longest running ``legacy" RV surveys that provided a long temporal baseline for many targets in the program that allowed for characterization of very long orbital period giant planets. Due to the nature of the AAPS program that is optimized for sampling long period orbits, the cadence of UCLES RVs is low and is thus not useful for detecting short period planets in this case. Nevertheless, HD~134606 was observed by UCLES 66 times from April 1998 to July 2015 for a total of 6,315 days of coverage. The data reduction procedure can be found in \citet{butler1996b}. Combined with the HARPS observations, an RV baseline of almost 7,000 days, or a little over 19 years in total was established that could potentially reveal the presence of previously undetected long period companions. We present a portion of the UCLES RVs in Table~\ref{tab:ucles}.

\begin{deluxetable}{lcr}[htbp]
    \tablecaption{UCLES RV Measurements of HD~134606.
    \label{tab:ucles}}
    \tablewidth{\columnwidth}
    \tablehead{
        \colhead{Time (BJD - 2,450,000)} &
        \colhead{RV (m~s$^{-1}$)} &
        \colhead{$\sigma$ (m~s$^{-1}$)}}
    \startdata
    917.23786 & 6.09 & 2.49 \\ 
    1,213.28251 & 12.04 & 2.36 \\ 
    1,274.27551	& -2.00 & 2.94  \\
    1,384.01069	& 4.75 & 2.07 \\ 
    1,683.03419	& 6.18 & 1.95 \\ 
    1,684.10581 & 8.54 & 2.18 \\ 
    1,718.08292 & 8.06 & 2.03 \\
    1,766.87982 & 12.81 & 2.07 \\
    1,767.92616 & 10.84 & 1.92 \\
    1,984.20376 & -4.10 & 2.52 \\\hline
    \enddata
    \tablecomments{Only a portion of the UCLES RVs are shown here. The full dataset can be accessed online in machine readable form.}
\end{deluxetable}

\subsection{Activity Indicators}
\label{sec:activityobs}
\subsubsection{HARPS}

A common problem many RV searches face is the confounding effect of stellar activity on the RV time series. Commonly, stellar rotation signals due to active regions or dark spots on the stellar surfaces may introduce RV semiamplitudes similar to those of small mass planets that would last for multiple stellar rotation periods \citep{robertson2015b, giles2017, robertson2020} whereas stellar magnetic cycles induced by the stellar dynamo effect could exhibit long-term periodic signals resembling induced RV by long-period gas giant planets that may last from several years to decades \citep{meunier2010, dumusque2011,costes2021}. There have been numerous cases where previously identified exoplanet candidates were later refuted because the RV signals were actually of stellar activity origins or at their harmonics \citep{robertson2014a,robertson2014b,robertson2015b,kane2016a,lubin2021,simpson2022}. It is therefore of the utmost importance that any RV observations should have at least one accompanying stellar activity indicator to disentangle stellar activity induced RV variations from those due to genuine exoplanet candidates.

The re-reduction of the HARPS RVs from the \citet{trifonov2020} HARPS RV database not only improves the RV precision, but also provides many activity indicators for each star that could be useful for singling out activity signals. In this work, we make use of all the available activity indicators from the database, namely, H$\alpha$ index, chromatic index (CRX), differential line width (dLW), Na I D index (NaD1), Na II D index (NaD2), cross correlation function (CCF) bisector inverse slope span (BIS), CCF full width at half maximum (FWHM), and CCF line contrast. The latter three activity indicators were derived by the HARPS Data Reduction Software (DRS) whereas the rest of the indicators were obtained from the re-reduction using the SERVAL pipeline. Briefly, H$\alpha$, NaD1, and NaD2 indexes measure emissions in these spectral lines that are known to be sensitive to activity. CRX provides information of the wavelength dependence of the RVs extracted from each echelle order and dLW measures the variations in line widths of spectral lines. Details regarding the indicators from the SERVAL spectral analysis can be found in \citet{zechmeister2018}.

Unfortunately the HARPS RVs provided by \citet{trifonov2020} do not include measurements of \shk index values for the spectra, which are known to correlate well with stellar chromospheric activity. In order to monitor stellar magnetic variability of HD~134606 via chromospheric emission in the Ca II H \& K lines, we compute the Mt.~Wilson \shk index \citep{vaughan1978,duncan1991} for the HARPS spectra. \shk is essentially a ratio of flux in the H \& K line cores to that in reference continuum bands on either side of the lines. We measure the calcium index using the 1D HARPS spectra as produced by the ESO data reduction pipeline\footnote{Based on data obtained from the ESO Science Archive Facility on Oct 25, 2022 with DOI: \url{https://doi.org/10.18727/archive/33}}. \shk values are computed using the \texttt{ACTIN 2} software package \citep{gsilva2018,gsilva2021}, and use the \texttt{pyrhk} extension to calibrate the index to the original Mt.~Wilson scale. The full \shk time series for the HARPS spectra contains a number of extreme outliers, all of which come from spectra with low signal-to-noise (SNR) ratios. We have thus limited our analysis to spectra with SNR $\geq~25$ in the spectral order containing the H \& K lines.

\subsubsection{UCLES}

The limited spectral coverage of the UCLES spectrograph makes it impossible to provide simultaneous coverage of the Iodine region for precise extraction of RVs from the stellar spectra and the Ca II H \& K region for derivation of stellar activity S-index measurement. In order to provide the RVs with an accompanying stellar activity indicator for the UCLES data, we measure the equivalent widths (EW) of H$\alpha$ absorption lines in the UCLES spectra to detect variations in the EW as a proxy to stellar activity. We follow the similar approach of the H$\alpha$ EW analysis described in \citet{robertson2014a} and \citet{wittenmyer2017a}, using an automated algorithm for normalizing the continuum and identifying telluric contamination close to the H$\alpha$ line. Although employing the H$\alpha$ line profile as a stellar activity indicator for a G-type star is non-ideal, it has been shown by \citet{silva2022} that the use of the H$\alpha$ line for activity studies can be optimized with a proper choice of bandwidth for extracting line profile changes. \citet{silva2022} pointed out that the usual choice of a broader 1.6\AA{} bandwith can sometimes lead to degradation of activity signals due to the inclusion of flux in the line wings, whereas a narrower 0.6\AA{} bandwidth for H$\alpha$ can circumvent the issue and maximize the correlation between H$\alpha$ and Ca II H\&K. We therefore adopt the recommendation of using a 0.6\AA{} bandwidth for calculating the H$\alpha$ EW to identify stellar activity signals in the UCLES data for HD~134606. 

\subsection{Speckle Imaging}
\label{sec:imagingobs}

In an effort to search for and/or rule out nearby stellar companions HD~134606, we carried out a speckle imaging observation using the Zorro instrument at Gemini South on July 24, 2021. Zorro is a dual-channel, dual-plate-scale imager that is able to observe simultaneously in two bands obtaining diffraction limited images with a field-of-view (FOV) of 6.7 arcseconds \citep{scott2021}. The simultaneous imaging is achieved by the use of a dichroic beamsplitter which splits collimated beam at the wavelength of 675 nm, with red channel light passed through and blue channel light reflected by it. The light in each channel is then recorded onto an electron-multiplying charged coupled device where two speckle patterns are simultaneously recorded using a sequence of 1000 short 60 ms exposures. These images were combined using Fourier analysis techniques, used to produce reconstructed speckle images, and examined for stellar companions \citep{horch2011,howell2011,scott2018}. Detailed instrument description, calibration, and data reduction process can be found in \citet{scott2021}. At the time of our observation, the blue channel was not used for speckle imaging due to instrument alignment issue and we only obtained data from the red channel with filter of central wavelength and bandwidth of 832~nm and 40~nm, respectively. 

\subsection{Astrometry}
\label{sec:astrometryobs}

We take our absolute astrometry for HD~134606 from the Hipparcos-Gaia Catalog of Accelerations, or HGCA \citep{brandt2018,brandt2021}. The HGCA cross-calibrates Hipparcos \citep{esa1997,vanleeuwen2007} and Gaia \citep{gaia2021} astrometry onto a common reference frame with recalibrated uncertainties suitable for orbit fitting. The HGCA provides a proper motion of HD~134606 from Hipparcos near Jyr 1991, a proper motion from Gaia near Jyr 2016, and a long-term proper motion given by the positional difference between Hipparcos and Gaia divided by their time baseline. Discrepancies between these three proper motions signify acceleration in an inertial reference frame.

\begin{deluxetable*}{lccc}
    \tablecaption{Absolute astrometry for HD~134606.
    \label{tab:hgca_astrometry}}
    \tablewidth{\columnwidth}
    \tablehead{
        \colhead{Parameter} &
        \colhead{Hipparcos} &
        \colhead{Hipparcos–Gaia} & 
        \colhead{Gaia EDR3}}
    \startdata
    $\mu_{\alpha*}$ (mas\,yr$^{-1}$) & $-177.80 \pm 0.45$ & $-177.908 \pm 0.015$ & $ -177.871 \pm 0.018$ \\
    $\mu_{\delta}$ (mas\,yr$^{-1}$) & $-165.14 \pm 0.59$ & $-164.653 \pm 0.017$ & $-164.709 \pm 0.021$\\
    ${\rm corr}(\mu_{\alpha*}, \mu_{\delta})$ & 0.28 & 0.25 & 0.20 \\
    $t_{\alpha}$ (Jyr) & 1991.23 & & 2016.27 \\
    $t_{\delta}$ (Jyr) & 1991.02 & & 2016.14
    \enddata
\end{deluxetable*}

The HGCA astrometry of HD~134606 suggests nonlinear motion, albeit at low significance, with a $\chi^2$ value of 8.4 for a model of constant proper motion. This is primarily due to a slight disagreement between the Gaia and the long-term proper motions, by far the most precise of the three measurements. Section \ref{sec:rv+ast} derives the implications of the absolute astrometry for the companions to HD~134606 and their orbits.

\subsection{Photometry}
\label{sec:transitobs}

\subsubsection{TESS}

We utilize time-series photometry obtained by the Transiting Exoplanet Survey Satellite \citep[TESS;][]{ricker2015} in order to search for the presence of transiting exoplanets and photometric stellar activity. HD~134606 was observed by TESS during Sectors 12, 38, and 39. We obtain 2-min cadence TESS light curve photometry from the Mikulski Archive for Space Telescopes (MAST). The light curve products include both the simple aperture photometry (SAP) and pre-search data conditioning simple aperture photometry (PDCSAP), which were processed by the Science Processing Operations Center (SPOC) pipeline \citep{jenkins2016}. We additionally remove any data that are flagged as being poor in quality or that are $>$~5$\sigma$ outliers relative to the root mean square (RMS) of the light curve. 

\subsubsection{ASAS-3}

Long term photometric monitoring allows identification of long period transiting planets as well as long period activity cycles. To check for any long term variation from photometry, we obtain the All Sky Automated Survey -- 3 (ASAS-3) V band data using the online catalogue portal. We pick data taken by aperture 3 given they yield the lowest scatter. Any data that are not of the best quality (quality grade A) are filtered out along with any outliers that are rejected after applying a 4$\sigma$ clip. The final ASAS-3 dataset results in 479 data points spanning from 2,452,441 to 2,455,111 BJD for a duration of 2,670 days, or about 7.3 years.


\section{Analysis}
\label{sec:analysis}

\subsection{Stellar Parameters}
\label{sec:star}

HD~134606 is a G6IV spectral type star and a previous publication on its age and radius indicate the possibility of it being near the end of its main sequence stage \citep{takeda2007}. However, the metal-rich nature of the star extends its main sequence lifetime to $\sim$12.5~Gyrs, as verified via isochrones from the MESA Isochrones \& Stellar Tracks (MIST) \citep{paxton2011,paxton2013,paxton2015,choi2016,dotter2016}. Here we derive new and updated stellar parameters for the HD~134606 system. We start by providing one of the high signal-to-noise HARPS spectra as an input to \texttt{SpecMatch-Emp} in order to constrain fundamental stellar parameters such as stellar effective temperature ($T_{\rm eff}$) and metallicity ($[\rm Fe/H]$). Details of how \texttt{SpecMatch-Emp} works can be found in \citet{yee2017}. In short, the provided spectrum is calibrated and matched against every other star available in the library and \texttt{SpecMatch-Emp} picks the five closest matches according to chi-squared statistics. A new best matching spectrum is then constructed using the five selected library spectra through linear combination. The set of coefficients of linear combination that minimized chi-squared when compared against the input spectrum is then chosen to output a weighted average of stellar parameters of the five selected stars along with their associated uncertainties. 

Derived $T_{\rm eff}$ and $[\rm Fe/H]$ values from \texttt{SpecMatch-Emp} are used as Gaussian priors for \texttt{EXOFASTv2}, which is used to derive a precise and consistent set of stellar parameters through modeling the spectral energy distribution of the star with archival broadband photometry from Gaia \citep{gaia2018}, Two Micron All Sky Survey \citep{skrutskie2006}, and Wide-field Infrared Survey Explorer \citep{wright2010} in combination with the aforementioned MIST stellar evolutionary models, all of which are included within 
\texttt{EXOFASTv2}. Additional constraints provided to \texttt{EXOFASTv2} include a Gaussian parallax prior from Gaia Data Release 3 (DR3) \citep{lindegren2021a, gaia2023} with a bias correction provided by \citet{lindegren2021b}, along with an upper limit on the V-band extinction using the galactic reddening maps from \citet{schlafly2011}. We provide derived stellar parameters from the converged \texttt{EXOFASTv2} fit in Table \ref{tab:star}.

\begin{deluxetable}{llr}[htbp!]
    \tablecaption{Derived stellar parameters for HD~134606.
    \label{tab:star}}
    \tablewidth{\columnwidth}
    \tablehead{
        \colhead{Parameters} &
        \colhead{Descriptions (Units)} &
        \colhead{Values}}
    \startdata
    $M_*$ & Mass ($M_{\sun}$) & $1.046^{+0.070}_{-0.059}$ \\ 
    $R_*$ & Radius ($R_{\sun}$) & $1.158^{+0.039}_{-0.036}$ \\
    $L_*$ & Luminosity ($L_{\sun}$) & $1.161^{+0.071}_{-0.049}$ \\
    $\rho_*$ & Density (cgs) & $0.95^{+0.13}_{-0.11}$ \\
    $\log{g}$ & Surface Gravity (cgs) & $4.330^{+0.044}_{-0.041}$ \\
    $T_{\rm eff}$ & Effective Temperature (K) & $5576^{+86}_{-85}$ \\
    $[\rm Fe/H]$ & Metallicity (dex) & $0.343^{+0.081}_{-0.084}$ \\
    $[\rm Fe/H]_{0}$ & Initial Metallicity & $0.343^{+0.074}_{-0.073}$ \\
    Age & Age (Gyr) & $7.3^{+3.6}_{-3.4}$ \\
    EEP & Equal Evolutionary Phase & $406^{+20}_{-41}$ \\
    $V$ mag & V-band magnitude & $6.854\pm0.010$ \\
    $A_{V}$ & V-Band Extinction (mag) & $0.079^{+0.084}_{-0.056}$ \\
    $\varpi$ & Parallax (mas) & $37.318 \pm 0.029$ \\
    $d$ & Distance (pc) & $26.797 \pm 0.021$ \\
    \enddata
    \tablecomments{Values reported are medians and 68\% confidence intervals for the stellar parameters. Initial metallicity is that of the star when it formed. Equal evolutionary phase (EEP) corresponds to static points in a star's evolutionary history. E.g., EEP of 202 and 454 represents zero and full age main sequence stages, respectively. For details see \citet{dotter2016} and \citet{eastman2019}. V band value taken from \citet{hog2000}.}
\end{deluxetable}

\subsection{Radial Velocity Solutions}
\label{sec:rvsol}

We utilize all of the available HARPS and UCLES data for our RV analysis. The joint dataset from the two sources provides an ideal opportunity to study both potential short and very long period signals at the same time thanks to the two long duration surveys contributing to a very long observation baseline as well as the high cadence observations conducted by HARPS. Special treatment is needed for both HARPS and UCLES data before the analysis. In June 2015, a new set of optical fibers was installed on HARPS, resulting a discontinuous jump in the RV time series \citep{locurto2015}. To account for the velocity offset caused by the fiber upgrade, we treat the pre-upgrade (HARPS1) and post-upgrade HARPS (HARPS2) data as two separate instruments. For UCLES, there appears to be an upward velocity offset in the RVs around 2,455,500 BJD. After analyzing the RV time series of other stars observed by UCLES in the AAPS program, the velocity shift at the same timestamp appears to be a common feature, suggesting the discontinuity in the RVs is probably not of astrophysical origin. However, there were no known upgrades or maintenance made to the spectrograph that would change the instrumental profile. The exact cause of the upward velocity shift in the RVs unfortunately is still unknown and is likely either instrumental or atmospheric. To err on the side of caution, we decide to treat UCLES data before and after 2,455,500 BJD as two separate instruments, UCLES1 and UCLES2, respectively.

\begin{figure}[htbp!]
  \includegraphics[trim=40 50 30 80,clip,width=\columnwidth]{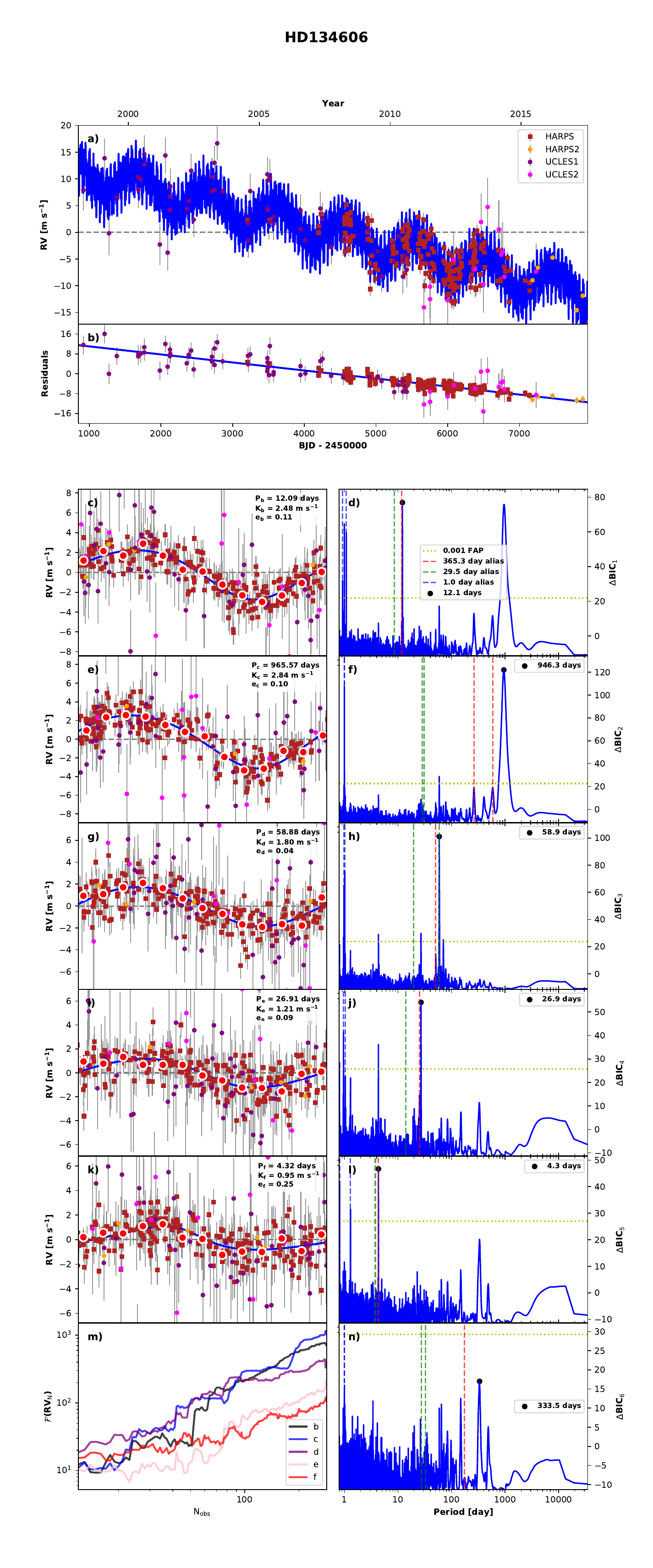}
  \caption{\texttt{RVSearch} result of the HD~134606 system. Upper panels show the best fit along with residuals to the model. Individual fit to each of the five signals are shown in the lower panels along with their periodograms. Panel (m) shows running periodogram for the five signals. No more significant signal is found in the residual as shown in panel (n).}
  \label{fig:rvsearch}
\end{figure}

We analysed four separate datasets from two instruments using \texttt{RVSearch} \citep{rosenthal2021}, an iterative planet searching tool to search for periodic signals within the combined RV dataset. \texttt{RVSearch} fits sinusoids to different fixed periods within the defined period search grid and utilizes the change in Bayesian Information Criterion ($\Delta$BIC) to measure of the goodness-of-fit of models. If there is a signal returned by \texttt{RVSearch} having a significance above certain empirical false alarm probability (FAP) threshold, which in our case we chose to be 0.1\%, then \texttt{RVSearch} carries out a maximum a posteriori (MAP) fit to refine the Keplerian model of that particular signal. If there are multiple signals in the datasets, $\Delta$BIC is once again used to determine if an n+1-planet model should be preferred over a previous n-planet model. This search algorithm has the advantage over the traditional Lomb-Scargle periodogram in that we are able to fit for instrumental velocity offsets, stellar jitters, linear or quadratic trend as part of the fitting process. In addition, already searched signals or known planet parameters are free to vary during the search and fitting process of additional signals, allowing the overall model to reach a higher maximum likelihood and therefore more precise orbital parameter results returned by the search \citep{rosenthal2021}. At each search iteration, the calculation of the FAP threshold follows a similar methodology as in \citet{howard2016} where a linear fit is modeled to the periodogram histogram in log scale against its $\Delta$BIC power values. The 0.1\% empirical FAP is then obtained by extrapolating the linear fit to a $\Delta$BIC value where only 0.1\% of the periodogram values are beyond this threshold. We set the algorithm to search between a minimum period of 0.8 days and a maximum period of five times the observing baseline of our combined data, which is around 35,000 days, while allowing linear and quadratic trend to be included in the search. The result of the planet search is shown in Figure \ref{fig:rvsearch}. The sub-panels are shown in the order of when each planet was detected by \texttt{RVSearch} rather than in order of orbital period. The search process successfully recovers two of the signals originally reported in M11: 12-day and 59-day, with FAP of each signal being 4.31$\times$10$^{-16}$ and 5.49$\times$10$^{-21}$, respectively. The third signal in M11 appears to manifest as a new signal with over double the originally reported period of $\sim$960 days (FAP = 4.57$\times$10$^{-27}$), suggesting the possibility that either M11 or our search algorithm here has found an alias rather than the true signal. Besides the three previously reported signals, the search detects two new low amplitude variations both with semiamplitude around 1 m~s$^{-1}$, yet with great significance: one at $\sim$27 days with FAP = 2.63$\times$10$^{-9}$ and the other at $\sim$4 days with FAP = 8.63$\times$10$^{-8}$. Some peaks appear to be significant around the 1-day periodicity but disappear after we fit for the strongest signal during each iteration, indicating those are 1-day aliases of the signals we recovered. On top of all recovered signals, a linear trend appears to be significant with a large amplitude, and is indicative of an additional long period signal associated with either a physical companion or a magnetic cycle. The significance of all returned signals show a monotonic upward trend according to the running periodogram in panel (m), especially during the HARPS observing period, implying that these signals are unlikely to arise from pure noise and are not only temporary and limited to specific seasons.

The MAP fit result for each returned signal from \texttt{RVSearch} is then passed to the RV modeling toolkit \texttt{RadVel} \citep{fulton2018a} as an initial guess to sample Keplerian model posteriors and estimate parameter uncertainties using Markov Chain Monte Carlo (MCMC). We use $P$, $T_{c}$, $\sqrt{e}$cos$\omega$, $\sqrt{e}$sin$\omega$, and $K$ as the fitting basis, where $P$, $T_{c}$, $e$, $\omega$, and $K$ are orbital period, time of inferior conjunction, eccentricity, argument of periastron, and RV semiamplitude, respectively. All parameters including linear trend, quadratic trend, instrumental jitters, and instrumental offsets are allowed to vary freely during the fitting process except for eccentricity and semiamplitude where the $e$ prior is forced to be less than 1 and $K$ is forced to be positive. Once again, we treat the UCLES and HARPS datasets before and after the velocity offsets as different instruments. The MCMC chain is carried out with 8 independent ensembles in parallel where each has 80 walkers that can take up to 15,000 steps. Chain convergence is evaluated under the following criteria: Gelman-Rubin statistic ($<$~1.01), minimum autocorrelation time factor ($>$~40), maximum relative change in autocorrelation time ($<$~0.03), and minimum independet draws ($>$~1000). The chain successfully converged before the walkers could reach the maximum allowed steps and we present the 16\%, 50\%, and 84\% solution along with the maximum likelihood result for both the fitted orbital parameters as well as derived physical parameters of the planetary candidates in Table \ref{tab:param}.

\begin{deluxetable*}{ccccccccc}[htbp!]
    \tablecaption{System parameters of the five planetary candidates in HD~134606.
    \label{tab:param}}
    \tablehead{
        \colhead{Data/Solution} & 
        \colhead{Planet} &
        \colhead{$P$ (days)} &
        \colhead{$e$} &
        \colhead{$\omega$ (deg)} &
        \colhead{$T_{p}$ (BJD - 2,455,000)} &
        \colhead{$K$ (m~s$^{-1}$)} &
        \colhead{$a$ (au)} &
        \colhead{$M_{p}\mathrm{sin}i$ ($M_{\rm \oplus}$)}
    }
    \startdata
    \multirow{5}{*}{\shortstack{HARPS1 Only \\ Quantiles}} 
     & b & $12.0897^{+0.0019}_{-0.0018}$ & $0.113^{+0.056}_{-0.057}$ & $150^{+30}_{-28}$ & $2.74^{+0.96}_{-0.95}$ & $2.5 \pm 0.13$ & $0.1046^{+0.0023}_{-0.0024}$ & $9.14^{+0.65}_{-0.63}$\\
     & c & $58.947^{+0.056}_{-0.054}$ & $0.053^{+0.058}_{-0.037}$ & $160^{+115}_{-92}$ & $26^{+16}_{-20}$ & $1.78 \pm 0.14$ & $0.3009^{+0.0065}_{-0.0069}$ & $11 \pm 1$\\
     & d & $958.7^{+6.3}_{-5.9}$ & $0.121^{+0.044}_{-0.05}$ & $191^{+25}_{-22}$ & $100^{+64}_{-59}$ & $2.83^{+0.13}_{-0.14}$ & $1.932^{+0.043}_{-0.045}$ & $44.5 \pm 2.9$\\
     & e & $4.31943^{+0.00075}_{-0.00068}$ & $0.22^{+0.14}_{-0.13}$ & $29^{+50}_{-63}$ & $500.03^{+0.59}_{-0.67}$ & $0.91^{+0.14}_{-0.13}$ & $0.0527^{+0.0011}_{-0.0012}$ & $2.31^{+0.36}_{-0.35}$\\
     & f & $26.9^{+0.019}_{-0.017}$ & $0.091^{+0.11}_{-0.066}$ & $65^{+48}_{-138}$ & $503.4 \pm 5.7$ & $1.16 \pm 0.14$ & $0.1784^{+0.0039}_{-0.0041}$ & $5.52^{+0.74}_{-0.73}$\\
    \hline
    \multirow{5}{*}{\shortstack{HARPS1 Only \\ Max Likelihood}} 
     & b & 12.0896 & 0.124 & 149 & 2.72 & 2.5 & 0.1019 & 8.43 \\
     & c & 58.945 & 0.016 & 126 & 27 & 1.78 & 0.293 & 11\\
     & d & 959.5 & 0.12 & 188 & 107 & 2.83 & 1.882 & 41.9\\
     & e & 4.31936 & 0.23 & 32 & 500.02 & 0.92 & 0.0513 & 2.12\\
     & f & 26.897 & 0.123 & 80 & 503.3 & 1.19 & 0.1737 & 5.72\\
     \hline
     \multirow{5}{*}{\shortstack{All Data \\ Quantiles$^{a}$}} 
     & b & $12.089^{+0.0016}_{-0.0015}$ & $0.092^{+0.054}_{-0.053}$ & $157^{+38}_{-34}$ 
     & $3.0 \pm 1.2$ & $2.47 \pm 0.13$ & $0.1046^{+0.0023}_{-0.0024}$ & $9.09^{+0.64}_{-0.63}$ \\
     & c & $58.883^{+0.041}_{-0.039}$ & $0.055^{+0.062}_{-0.04}$ & $172^{+80}_{-86}$ & $31.6^{+9.9}_{-21}$ & $1.81 \pm 0.14$ & $0.3007^{+0.0066}_{-0.0069}$ & $11.31^{+1.0}_{-0.99}$ \\
     & d & $966.5^{+5.3}_{-6.9}$ & $0.092 \pm 0.045$ & $194^{+34}_{-28}$ & $100^{+84}_{-79}$ 
     & $2.83 \pm 0.13$ & $1.941^{+0.043}_{-0.046}$ & $44.8 \pm 2.9$ \\
     & e & $4.3203^{+0.00051}_{-0.00047}$ & $0.2^{+0.14}_{-0.13}$ & $54^{+45}_{-57}$ & $500.27^{+0.56}_{-0.61}$ & $0.92 \pm 0.13$ & $0.0527^{+0.0011}_{-0.0012}$ & $2.34^{+0.35}_{-0,34}$ \\
     & f & $26.915 \pm 0.016$ & $0.081^{+0.1}_{-0.059}$ & $63^{+63}_{-166}$ & $504.4^{+7.3}_{-6.7}$ & $1.18 \pm 0.14$ & $0.1784^{+0.0039}_{-0.0041}$ & $5.63^{+0.72}_{-0.69}$ \\
     \hline
     \multirow{5}{*}{\shortstack{All Data \\ Max Likelihood}} 
     & b & 12.089 & 0.1 & 154 & 3.0 & 2.48 & 0.1009 & 8.12 \\
     & c & 58.88 & 0.05 & 166 & 35 & 1.81 & 0.2898 & 10.62 \\
     & d & 967.9 & 0.092 & 200 & 117 & 2.86 & 1.872 & 39.9 \\
     & e & 4.32016 & 0.26 & 57 & 500.25 & 0.94 & 0.0508 & 2.22 \\
     & f & 26.912 & 0.06 & 115 & 505 & 1.21 & 0.172 & 4.8 \\
    \enddata
    \tablecomments{Model based on the entire UCLES and HARPS data measures a linear trend of $-0.00340 \pm 0.00020$ m~s$^{-1}$~d$^{-1}$ with the maximum likelihood value of -0.0034 m~s$^{-1}$~d$^{-1}$. Linear trend measurement from the model based on the HARPS1 data alone provides a consistent value of $-0.00333 \pm 0.00018$ m~s$^{-1}$~d$^{-1}$ with the maximum likelihood value of -0.00335 m~s$^{-1}$~d$^{-1}$. $\omega$ values in each row of the table are those of the star, not of the planets.}
    \tablenotetext{a}{The best-fit model we employ for the rest of this work.}
\end{deluxetable*}

UCLES data are extremely helpful in providing a longer observation baseline. However, it is worth noting that due to the limited precision of UCLES, the five signals recovered in Figure \ref{fig:rvsearch} are completely driven by the HARPS data. When running \texttt{RVSearch} on the UCLES data alone, no signal can be returned other than the linear trend. This is expected considering UCLES data exhibit lower precision and greater scatter: mean error of 1.60 and 1.74 m~s$^{-1}$ for UCLES1 and UCLES2, respectively, compared to 0.72 and 0.70 m~s$^{-1}$ for HARPS1 and HARPS2. In addition, UCLES suffers significantly higher instrumental jitters than the HARPS data, with jitters for UCLES1 around 2.98 m~s$^{-1}$ and UCLES2 around 5.20 m~s$^{-1}$, whereas those for HARPS1 and HARPS2 are about 1.80 and 0.70 m~s$^{-1}$, respectively. For that reason, we carry out another fitting process using only the HARPS1 data. HARPS2 data are not combined with HARPS1 data in this case to simplify the fitting since we are able to fit with two fewer free parameters at a cost of only five fewer data points. The MCMC chain converges quickly and we provide solutions of the 5-planet candidate model using the HARPS1 data alone in addition to those from the combined dataset including all the UCLES and HARPS data in Table \ref{tab:param}. The RV solutions and uncertainties from models utilizing all the data and only HARPS1 data are consistent with each other. Taking into consideration of similar model result but with much longer observation baseline, we opt for the solution with all the data as the model we employ for the rest of this work. Model comparison based on the Akaike Information Criterion (AIC) shows all five planetary candidate signals are favored in the final solution along with a linear trend, which bears a significance of about 17$\sigma$. However, there is no great statistical evidence for a quadratic trend in either of the solutions. Models with fewer free parameters containing planets less than five all have $\Delta$AIC $>$~47, indicating these models are not supported by our data and are completed ruled out.

\subsection{RV Injection-Recovery Test}
\label{sec:rvinjection}

\begin{figure*}[htbp]
    \begin{center}
        \begin{tabular}{cc}
            \includegraphics[trim=40 40 10 20,clip,width=0.5\textwidth]{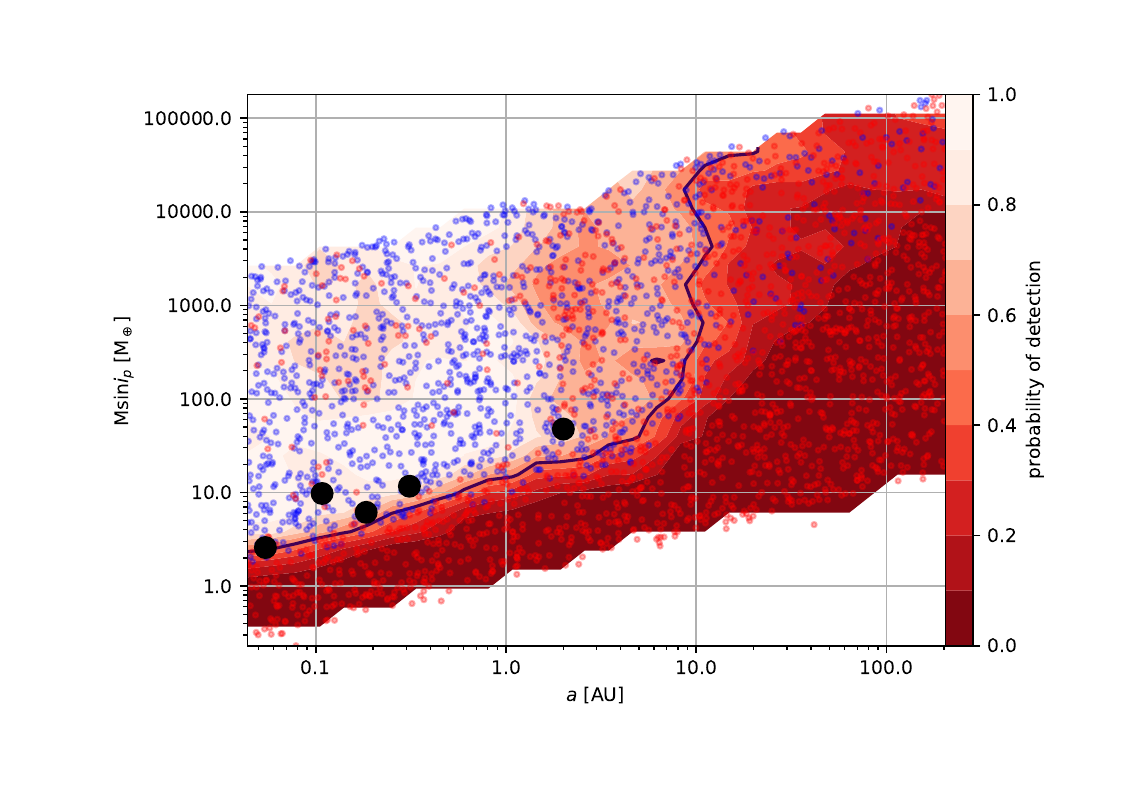} &
            \includegraphics[trim=40 40 10 20,clip,width=0.5\textwidth]{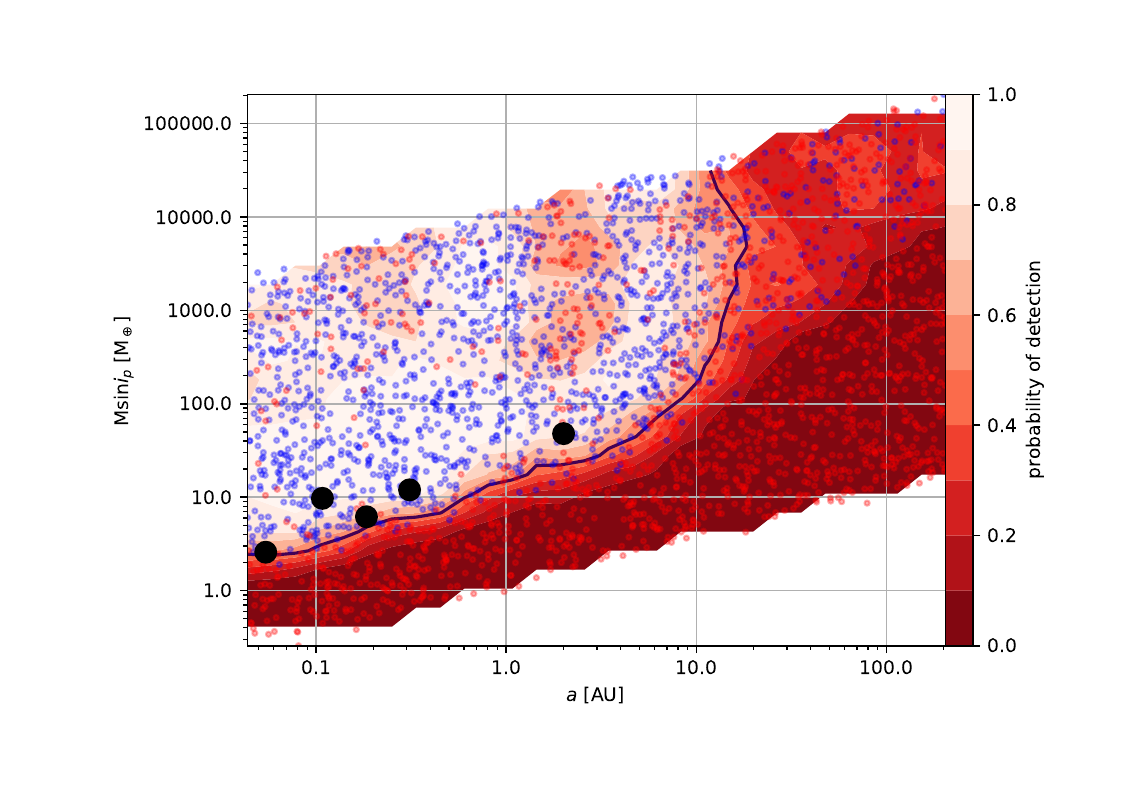} \\
        \end{tabular}
    \end{center}
    \caption{RV completeness contour of HD~134606 from the injection-recovery tests. Tests conducted with only HARPS1 are shown in the left panel and those carried out with all the RV data are in the right panel. Blue dots are injected signals that were successfully recovered whereas red dots were not. The black line represents the 50\% detection probability, with redder color showing lower probability.}
    \label{fig:rvinjection}
\end{figure*}

The sensitivity of RV data gives information as to the possible presence of additional companions that may lurk in the system below the current detection threshold. Such an analysis is an important component of a complete system characterization \citep{wittenmyer2020a, dalba2021, li2021}. Here, we characterize sensitivity of our RV data by carrying out injection-recovery tests, where we inject synthetic signals into the RV data and attempt to recover these injected signals using \texttt{RVSearch}. We inject 3,000 synthetic planets with orbital period and minimum mass drawn from log-uniform distributions and orbital eccentricity from the beta distribution with parameters from \citet{kipping2013b}. The result of the injection-recovery tests are then used to compute the search completeness contour in the minimum mass versus semimajor axis space. As before, due to the higher precision HARPS data have, we carry out a separate test with only the HARPS1 data in addition to the one with all the HARPS and UCLES RVs and we show the results as two panels in Figure \ref{fig:rvinjection}.

The five planetary candidates are shown in solid black dots and are all above the 50\% completeness curve, with the candidate planet e at 4.32 days sitting closest to the curve given it has the lowest semiamplitude among all signals and being closest to the noise floor of our data. Both panels show a similar level of sensitivity to low amplitude signals thanks to the presence of HARPS data. The combined RV (right panel) show higher sensitivity to higher mass companions at longer periods compared to the HARPS only data (left panel) due to the longer baseline of UCLES data and their sampling optimized for long period companion search. In both cases, potential planets with similar or larger masses than the five in Table \ref{tab:param} appear to be recoverable from our data within 2~au but were not detected, indicating no more such planetary presence in the inner region of the system. According to the right panel, presence of a gas giant planet is very unlikely within 10~au orbital radius and the ability to recover very long period companions is fairly limited beyond the baseline of the combined dataset ($\sim$19 years, or $a$$\sim$7~au), where the detection probability drops below 50\% for a Jupiter-mass planet at a little over 10~au. Therefore, the possibility of one very long orbital period companion hinted by our RV linear trend cannot be ruled out.

\subsection{Speckle Imaging Constraint}
\label{sec:imagingsol}

Reconstructed images derived from the speckle imaging reduction process mentioned in section \ref{sec:imagingobs} are used to examine for presence of stellar companions around HD~134606. Similar to the process presented in \citet{horch2011} and \citet{howell2011}, we estimate the location of a secondary peak under the assumption of the presence of a companion and modeled the fringe pattern using the position of the secondary. The signal of the tentative companion is then inspected by various metrics in the procedure and also manually to check whether the secondary peak is due to a genuine star or is because of a noise spike or cosmic ray. Only a true companion detection would be retained. For HD~134606, no companion is detected in our speckle imaging data using the red channel of the instrument, suggesting that either there is no companion within the FOV of the instrument, or maybe a companion is simply not bright enough to be detectable. In this case, it is useful to derive a relative detection limit to answer the question of how bright a potential companion could be while remaining below the detection sensitivity of the instrument.

\begin{figure}[htbp!]
  \includegraphics[trim=5 5 5 17,clip,width=\columnwidth]{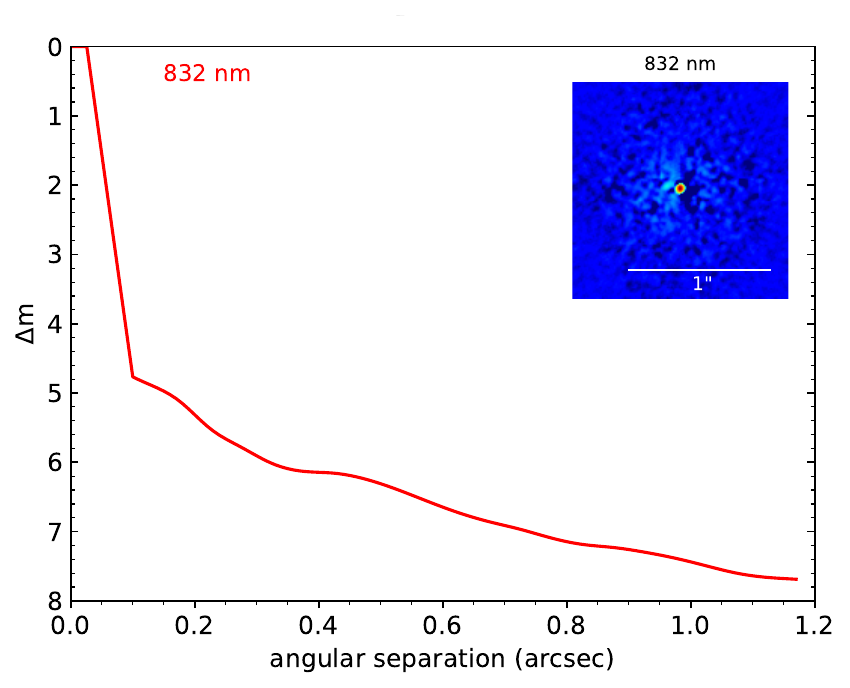}
  \caption{Speckle imaging detection limit in terms of instrumental magnitude difference as a function of separation from the primary star. No bright companion was detected. Only the red channel was used for observation as the blue channel suffered an instrument alignment issue.}
  \label{fig:speckle}
\end{figure}

To get a sense of the limiting magnitudes of our imaging data as a function of separation from the host star, we examine the distribution of all local maxima and minima in the background of the image by drawing five concentric annuli centered on the primary star, each with width of 0.2\arcsec{} centered at radii of 0.2, 0.4, 0.6, 0.8, and 1.0 arcsec. We treat each maximum as a potential stellar signal and compute the mean peak values within each annulus and the corresponding magnitude difference. Since the distributions of maxima and minima are essentially mirror images of each other with similar absolute mean and standard deviation values, we average the values from both maxima and minima when calculating the magnitude difference. The detection limit as a function of separation is then estimated using a conservative 5$\sigma$ limit, which we define as being 5 times the standard deviation of the mean peak values plus the mean peak values in each annulus. The detection limit derived for each annulus in terms of instrumental magnitude difference ($\Delta$m) is used as a conservative upper limit above which stars should be detected, thus providing a constraint of the possible undetected low mass companions nearby. We show the detection limit curve of our speckle imaging observation from the red channel of the instrument in Figure \ref{fig:speckle}.

\begin{figure}[htbp!]
  \includegraphics[trim=10 5 30 10,clip,width=\columnwidth]{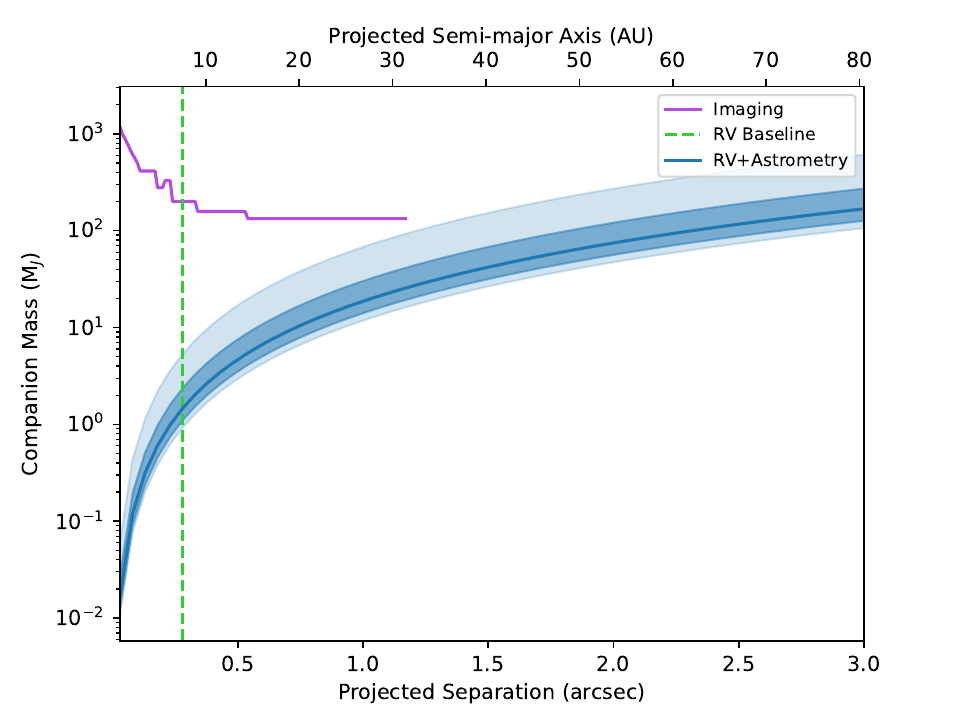}
  \caption{Detection sensitivity curves from speckle imaging and joint RV+astrometry in terms of Jupiter masses as a function of separation from the central star. The purple sensitivity curve for speckle imaging is an upper limit, while the blue curve is a measurement for joint RV and astrometry constraint. RV observational baseline is indicated by the vertical green dashed line. Any potential companion is likely to be either a high mass giant planet, brown dwarf, or a very low mass star.}
  \label{fig:limits}
\end{figure}

The derived 5$\sigma$ $\Delta$m upper limit is then used to place upper limits on masses of potential companions as a function of separation similar to what has been done in previous works \citep{kane2014c,kane2019b,dalba2021}. We first iterate through all the available stellar spectral types in the Pickles spectral library \citep{pickles1998} and calculate respective luminosity ratios between the primary and the companion, and therefore, estimate instrumental magnitude differences from the given spectral type of the companion ($\Delta$m$^\prime$) with the Zorro 832~nm red band filter. The calculated $\Delta$m$^\prime$ were then compared to the $\Delta$m limits we derived from our speckle imaging observation at each separation and we pick the spectral type that yields the closest matching magnitude difference as the star that would yield the magnitude upper limit at that separation. For the selected spectral type of the companion at each location, we once again compute luminosity ratios and magnitude differences, this time in the apparent V-band magnitude ($\Delta$m$_\mathrm{V}$). Using the known V-band magnitude of the primary star and the distance to the system, calculated absolute magnitude values M$_\mathrm{V}$ are then used along with the mass-luminosity relations \citep{henry1993} to yield the final upper companion mass limit as a function of angular separation using the speckle imaging data. The derived mass upper limit curve is shown in Figure \ref{fig:limits}.

\subsection{Joint RV+Astrometry Constraints} 
\label{sec:rv+ast}

HD~134606 has a Renormalised Unit Weight Error value around 1.022 from Gaia DR3, suggesting the single-star model is a good fit to the astrometric observations. However, such a binarity flag is only sensitive to binaries that contribute a significant fraction of the light and/or are on a period shorter than a few times the Gaia mission life. Therefore, additional large mass companions are possible at larger orbital separations even if Gaia indicates no binarity for the star. HD~134606 has a detectable RV acceleration and a detectable astrometric acceleration between Hipparcos and Gaia, though the astrometric acceleration is less than 3$\sigma$ significant. The RVs themselves show no deviation from a linear trend over 19 years of monitoring (see section \ref{sec:rvsol}). For the following analysis, we therefore assume that the star's acceleration is constant over the entire time baseline monitored by UCLES, HARPS, Hipparcos, and Gaia.  

The Hipparcos-Gaia Catalog of Accelerations includes HD~134606 as a low-significance astrometric accelerator; its $\chi^2$ value with respect to constant proper motion is 8.4 with two degrees of freedom (2.4$\sigma$ Gaussian equivalent significance). We compute the difference in proper motion between Gaia and the Hipparcos-Gaia mean proper motion separately in R.A.~and Decl. We divide this difference by one-half the time baseline between Hipparcos and Gaia, i.e., by 12.43 years for R.A.~and by 12.45 years for Decl. We add the proper motions in R.A.~and Decl.~in quadrature and use standard propagation of uncertainties to obtain an astrometric acceleration of $5.4 \pm 1.9$~$\mu$as\,yr$^{-2}$. Finally, we use the measured Gaia parallax of $19.14 \pm 0.08$~mas to convert this proper motion acceleration to a physical value of $0.68 \pm 0.24$~m\,s$^{-1}$yr$^{-1}$.

The accelerations in astrometry and in RV together constrain the ratio $M_{\rm sec}/\rho^2$, where $\rho$ is the projected separation. We define $r$ to be the physical separation of the star and its companion and $\phi$ to be the angle between this separation vector and the plane of the sky. We then combine 
\begin{align}
    \label{eqn:acc}
    a_{\rm RV} &= \frac{G M_{\rm sec}}{r^2} \sin \phi \\
    a_{\rm ast} &= \frac{G M_{\rm sec}}{r^2} \cos \phi \\
    \rho &= r \cos \phi
\end{align}
where $a_{\rm RV}$ and $a_{\rm ast}$ are the accelerations along and perpendicular to the line of sight, respectively, to obtain
\begin{equation}
\frac{GM_{\rm sec}}{\rho^2} = \left(1+\frac{a_{\rm RV}^2}{a_{\rm ast}^2} \right) \sqrt{a_{\rm RV}^2 + a_{\rm ast}^2} .
\end{equation}
We do not use standard propagation of uncertainties because the signal-to-noise ratio on the astrometric acceleration is modest and the true distribution for $GM_{\rm sec}/\rho^2$ will be significantly non-Gaussian. Instead, we use Monte Carlo with Gaussian uncertainties on $a_{\rm RV}$ and $a_{\rm ast}$ to obtain 
\begin{equation}
    \frac{M_{\rm sec}}{\rho^2} = 19_{-5}^{+11}\,\frac{M_{\rm 
 Jup}}{{\rm arcsec}^2}
\end{equation}
(at 16/50/84\% quantiles), or a 95\% confidence interval of 12--68\,$M_{\rm Jup}/{\rm arcsec}^2$. These constraints are shown as shaded intervals in Figure \ref{fig:limits}.

\subsection{Photometric Variability}
\label{sec:transitsol}

We search for photometric variability in the 2-min TESS light curves following the variability analysis of \citet{fetherolf2023}, which we describe briefly here. Variability is searched on short-timescales ($<$~13\,days) in the TESS light curves using a Lomb-Scargle periodogram \citep{lomb1976, scargle1982} and auto-correlation function. The Lomb-Scargle periodogram searched for periodic variability on timescales of 0.01--1.5\,days and 1--13\,days in each individual TESS Sector (12, 38, and 39) in both the SAP and PDCSAP photometry. Photometric variations in the SAP photometry are consistent with known timings of spacecraft thruster firing events, and these variations have been removed in the PDCSAP photometry. The PDCSAP photometry is consistent with being flat (normalized LS power $<$~0.01), with an RMS of 235, 180, and 190\,ppm in Sectors 12, 38, and 39, respectively. Therefore, we do not detect photometric variability in the TESS light curves. We also search for the presence of transiting planets by investigating the phase folding the concatenated TESS light-curve on the measured radial-velocity orbital periods for each planet in the HD~134606 system. However, we do not detect any evidence for transit events in the TESS light curves at a limit of 50\,ppm. 

We carry out a similar search for the ASAS-3 photometry. The scatter and precision of the ASAS-3 data are unfortunately too poor for transit detection of our planetary candidates here and are therefore not useful for this purpose. However, running the data through a Generalized Lomb-Scargle (GLS) periodogram \citep{zechmeister2009} does reveal signatures with $\sim$1200-day and $\sim$4600-day periods within the ASAS-3 photometry that may indicate a long term stellar activity variation, which we will discuss in section \ref{sec:activitysol}.


\section{Stellar Activity}
\label{sec:activitysol}

An essential step of all RV analyses as mentioned in section \ref{sec:activityobs} is to check whether any of the signals recovered from the studied RV time series are susceptible to contamination by stellar activity, as signals such as stellar rotation, magnetic cycles and their aliases are often misidentified as exoplanet candidates. Here, we use activity indicators based on activity-sensitive spectral lines from both the HARPS and UCLES spectra as well as photometry from TESS and ASAS to identify signals of stellar activity origin.

\subsection{HARPS}

\begin{figure}[htbp!]
  \includegraphics[trim=0 0 0 0,clip,width=\columnwidth]{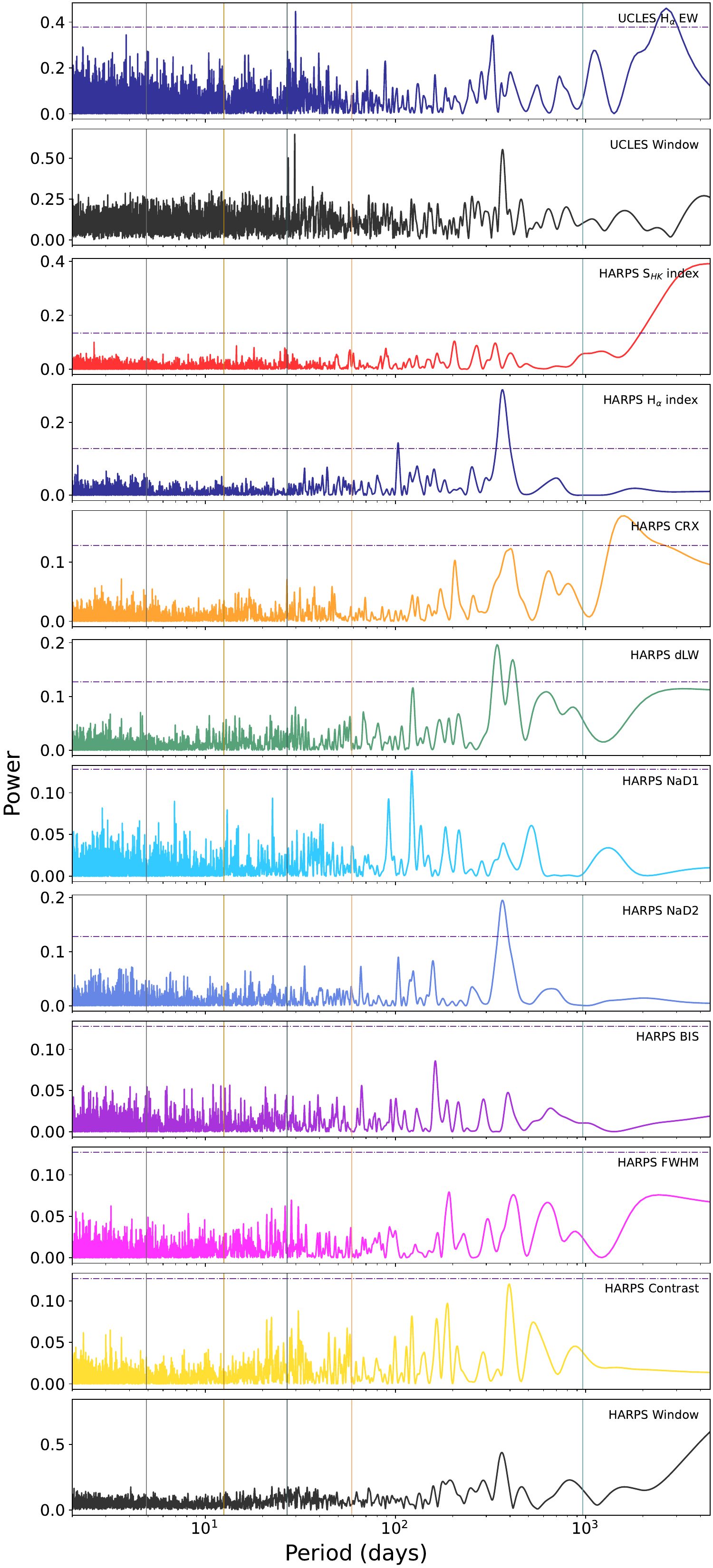}
  \caption{GLS periodogram of activity indicators for HARPS and UCLES data. Window functions for each are in black. Horizontal dashed lines are 0.1\% FAP and vertical solid lines represent period locations for the five planet candidates. No peaks in this periodogram coincide with any of the candidates' periods.}
  \label{fig:allactivity}
\end{figure}

We take the activity indicators from the HARPS database that includes all re-reduced RVs and their associated activity indicators, from both the SERVAL pipeline and ESO's default DRS pipeline \citep{trifonov2020}. Provided HARPS indicators include H$\alpha$ index, CRX, dLW, NaD1 index, NaD2 index, BIS, FWHM, and CCF line contrast. Details of these indicators can be found in section \ref{sec:activityobs} of \citet{zechmeister2018}. In addition, we derived our own \shk index from the HARPS re-reduced spectra and calibrated to the Mt.~Wilson scale (see section \ref{sec:activityobs}). The Ca II H\&K measurement later became available as part of the second version of HARPS RVBank \citep{perdelwitz2023} and we use it to double check our result from the \shk. We once again only utilize HARPS1 activity data since we do not want to introduce an unwanted vertical offset between data coming from before and after the fiber upgrade, especially considering there are only 5 data points for HARPS2. We detrend activity indicators that show a significant trend (see section \ref{sec:nature}) before passing them through a GLS periodogram to uncover any periodic features. Search range is limited between 2 and 4,500 days and we show the result in Figure \ref{fig:allactivity}. For each indicator, we mark the 0.1\% FAP threshold as indicated by the horizontal dashed line. Signals with powers extending above these horizontal lines have high significance and less than 0.1\% FAP. Additionally, the five planetary candidate signals we recovered in section \ref{sec:rvsol} are overlaid on each periodogram for reference. At first glance, there appears to be a peak occurring at 1-year periodicity in many of the indicators such as the H$\alpha$ index, CRX, dLW, NaD2, and CCF contrast, of which the H$\alpha$, dLW, and NaD2 data provide great signal power. This is likely due to the consistent sampling during each observing season when the target was visible. At longer periods, \shk index, CRX, and dLW all have strong peaks below or close to 0.1\% FAP, probably due to a stellar magnetic cycle. These peaks occur at periods comparable or longer than the HARPS1 baseline, which is just less than 4,000 days. Results of Ca II H\&K from \citet{perdelwitz2023} and our own \shk are consistent with each other, except the long-period weak exhibit a much weaker power, possibly due to different methods for flux extraction. One more activity signal shows up at a little over a 100-day period and is visible in H$\alpha$, dLW, and NaD1. The spectral window function \citep{dawson2010}, which is useful in identifying artificial periodicities incurred by the sampling frequency, displays no strong peaks for HARPS other than the 1-year signal, confirming the origin of it being the consistent yearly observations. The window function power rises towards longer periods are due to the testing frequency range extending beyond the HARPS observation baseline, which is a common behavior. No other significant HARPS activity peaks are found and none of the strong signals in the periodogram coincide with any of the five planetary signals.

\subsection{UCLES}

For the UCLES data, \shk index was unobtainable due to the limited coverage of the UCLES spectrograph. We therefore utilize the H$\alpha$ EW we derived with a 0.6\AA{} bandwidth to best use H$\alpha$ as a proxy to S-index for this G-type star. As mentioned in section \ref{sec:rvsol}, there appears to be a systematic velocity offset near 2,455,500 BJD in the RV time series. We therefore split the EW dataset into before and after the offset, the same way we treated the RVs, to avoid the offset leading to a spurious activity signal. We follow the same procedure as for the HARPS activity indicators and we show the pre-offset UCLES EW periodogram and spectral window function in Figure \ref{fig:allactivity} with a baseline around 4,500 days along with other HARPS indicators. Two peak stand out with less than 0.1\% FAP in the UCLES EW periodogram. The peak at $\sim$3,000-day periodicity appear to be indicative of the long-term magnetic cycle that some of the HARPS activity indicators detected, confirming its identity. The other peak detected by H$\alpha$ EW at around 29 days, which is also evident in the UCLES window function, is likely due to the observing sampling frequency affected by the moon's orbital period. Additionally, the 365-day periodicity that is strong in many of the HARPS indicators is also apparent. Activity periodicity search for the UCLES post-offset H$\alpha$ EW dataset reveals a negative result and all peaks detected in the pre-offset activity data do no overlap with the five candidate signals. 

It is worth mentioning the choice of using a 0.6 \AA{} bandwidth recommended by \citet{silva2022} for deriving our H$\alpha$ EW time series values indeed yield better result. The signals at both 29-day and 3,000-day display stronger power in the periodogram with 0.6\AA{} bandwidth data compared to 1\AA{} data. With the use of a traditional 1.6\AA{} bandwidth, the two signals are not even recovered. In this case, the use of a wider bandwidth for extracting the H$\alpha$ data provides a poor result, likely due to additional flux from the line wings degrading the signal as suggested by \citet{silva2022}. The use of a 0.6\AA{} bandwidth works well here, but it remains to be seen whether all studies should employ this strategy or its usage should be determined case by case.

\subsection{Photometry}

\begin{figure}[htbp!]
  \includegraphics[trim=0 0 0 0,clip,width=\columnwidth]{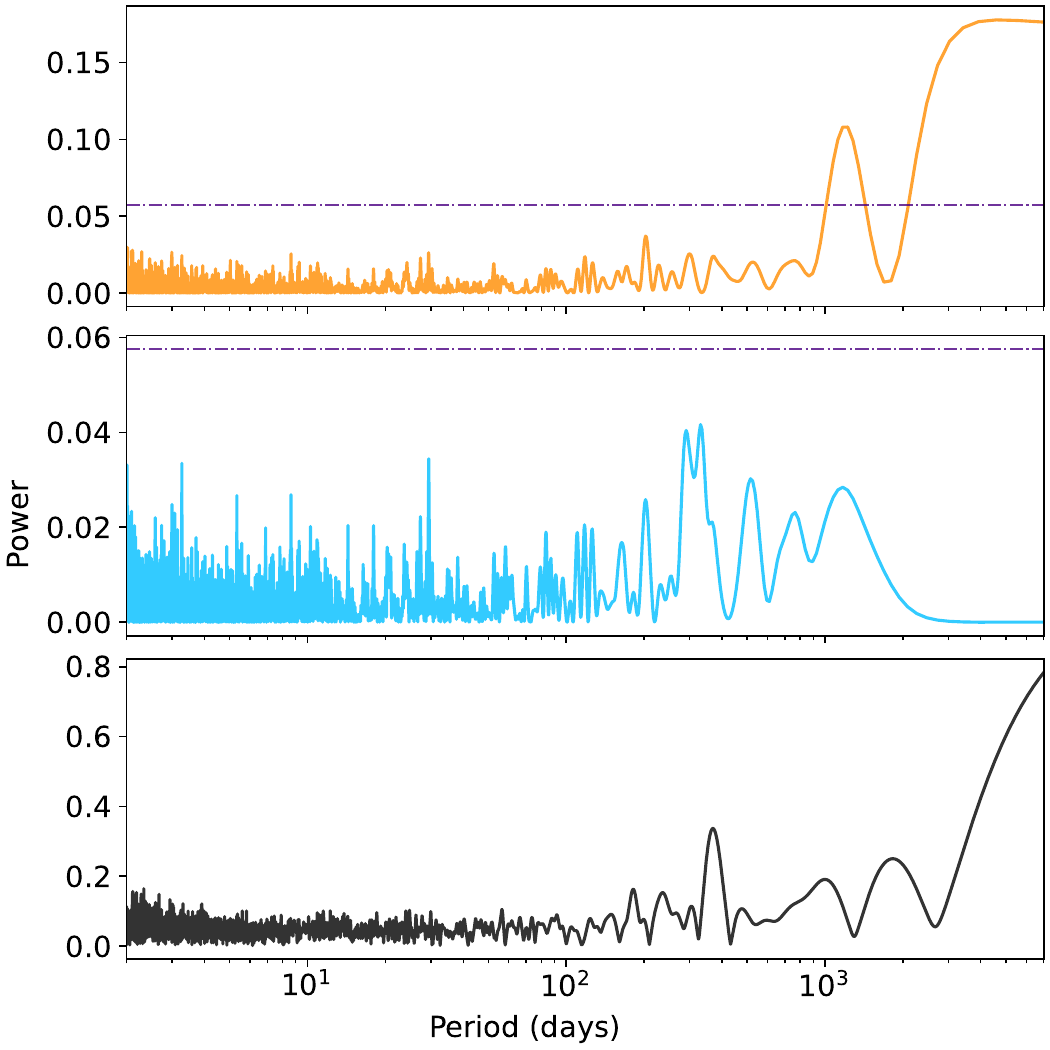}
  \caption{GLS periodogram of ASAS-3 photometry. Horizontal dashed lines are 0.1\% FAP thresholds. Upper and middle panel shows the periodogram before and after the subtraction of the 4,600-day peak, respectively. Lower panel shows the window function of the photometry.}
  \label{fig:asasactivity}
\end{figure}

Photometry is often useful as another diagnosis of stellar activity by looking at the photometric light curves and identifying variability \citep{meunier2019b, fetherolf2023, simpson2022}. We study time series photometry from TESS and ASAS-3. TESS analysis in section \ref{sec:transitsol} indicates no obvious variability present in the light curve and therefore no stellar rotation or other activity signals were detected in TESS photometry. The ASAS-3 data, however, appears to exhibit a long term cycle. Two peaks are detected, one around 1,200-day periodicity and the other at the upper end of the observing baseline. The higher peak has a longer period extending beyond the baseline of the ASAS-3 photometry and therefore the period cannot be precisely located. However, by extending the search range to longer periods allows us to broadly estimate the period of this signal of $\sim$4,600 days (Upper panel of Figure \ref{fig:asasactivity}). Subtracting this signal removes the other peak at 1,200-day as well, suggesting the 1,200-day signal might be an alias of the longer period signal. Although the location of the higher peak is poorly defined, the origin of it could very likely be the same magnetic cycle that is detected in other spectroscopic activity indicators. Although the window function power rises again towards longer periods, this is again the same window function behavior as we saw earlier for the HARPS data where the testing range is going beyond the observing baseline. 

\subsection{Rotation Period}

Neither spectral activity indicators nor photometry show any periodicities at short periods, indicating no short term activity signals persisting over the duration of the long observing baseline we have. Rotationally-modulated signals from starspots, however, are not coherent, and degrade over a few stellar rotation periods \citep{robertson2015b,robertson2020,giles2017,lubin2021}. Therefore, the power in rotation signals are typically lost over a long period of time if spots do not consistently occur on the stellar surface. To double check for the presence of a stellar rotation signal, we divide the HARPS1 dataset into seasons and attempt to retrieve any short-term periodic signatures. UCLES data are not utilized here since their sparse sampling were more optimized for the long-period planet search. We follow a similar procedure as before and use \texttt{GLS} to search for periodicities within the range of each season, $\sim$200-250 days, for each one of the HARPS activity indicators as well as for ASAS photometry. None of the periodograms return significant peaks and therefore we conclude the stellar rotation signal was not detectable in our spectral and photometric data, possibly due to the reduced level of activity at a relatively senior stage the star is at in its lifetime right now.

Given the non-detection of stellar rotation in our data, we attempt to estimate the rotation period of HD~134606 using the \texttt{pyrhk} function as a final check for stellar activity and possible aliases. The estimation calculates the chromospheric rotation period using activity-rotation relations from \citet{mamajek2008}, where the Rossby number is fitted against \logrphk for a sample of 169 solar-type stars. Rotation period is then derived from the Rossby number using convective turnover time relation as a function of B-V color \citep{noyes1984b}. For HD~134606, the time-averaged \logrphk is -5.1, a fairly quiet star, and the stellar rotation period is estimated to be $42.0 \pm 3.9$ days. This period and its aliases do not overlap with any of the five planetary candidates in Table \ref{tab:param}. However, given \logrphk of the inactive stars (\logrphk~$<$~-5.0) in the sample from \citet{mamajek2008} is only very weakly correlated with Rossby number, the derived rotation period should not be taken with trust and is therefore only used for a precautionary check for RV false positives. 


\section{Discussion}
\label{sec:discussion}

\subsection{Nature of the Recovered Signals}
\label{sec:nature}

Although the stellar activity checks we performed in section \ref{sec:activitysol} do not yield any significant periodicities that coincide with any one of the five recovered signals from Table \ref{tab:param}, there remains some questions regarding the validity of the five planets: 1) there are two seasons within HARPS CCF contrast or FWHM data that show a weak $\sim$30-day periodicity, indicating the need to further scrutinize our planet f due to its proximity to the lunar cycle period despite the activity signals having FAP greater than 0.1\% in the periodograms; 2) M11 reported a period for planet d that is nearly half of what we have recovered, which suggests that either M11 or this work is seeing an alias of the actual signal; 3) planets b, c, e, and f appear to be near integer ratios of each other, which could potentially be a sign of aliasing; 4) there is a significant linear trend present in the HARPS dLW, CCF contrast, and FWHM data, therefore making the physical companion nature of the RV linear trend questionable. Given these concerns, we perform final checks within the dataset to further verify the significance or nature of our recovered planet signals. Since all five planetary signals are driven by the HARPS data, we utilize only the HARPS1 data here to simplify the checking procedure.

\subsubsection{Planet f's 26.9-day Signal}

Planet f's 26.9-day period is close to the 29-day lunar cycle. To make it more worrisome, the period almost overlaps with the 1-year alias of the lunar cycle, making its planetary nature highly suspicious. Given that only two of the spectral activity indicators show a weak signal that may be attributed to the lunar cycle in two of the seasons (from BJD 2,456,311.9 to 2,456,516.5, and from BJD 2,454,879.89 to BJD 2,455,025.57), we create a synthetic dataset consisting of white noise with timestamps and error bars for the HAPRS1 dataset. The velocity scatter is drawn from a Gaussian distribution with standard deviation being the quadrature sum of original data point errors and RMS of the five-planet fit from section \ref{sec:rvsol}. Then, we inject a 29-day signal into the seasons in question with similar error and scatter. A periodogram analysis is then run with \texttt{RVSearch} on the entire synthetic data set and we find the injected signal is not recoverable with $<$~0.1\% FAP unless the variation amplitude is increased to $\sim$5 m~s$^{-1}$. In all scenarios, there is no indication of an 1-year alias of the lunar cycle being produced without the 29-day signal being recovered first, if that 29-day cycle is recoverable at all. This suggests that the time sampling and the lunar cycle within the season in question is extremely unlikely to have produced the 26.9-day period. As an additional check to see if such a periodicity is localized, we remove all the RVs within these two seasons and run another periodogram search. In this case, all five planets are once again recovered with FAP being 1.19$\times$10$^{-12}$, 8.26$\times$10$^{-9}$, 1.10$\times$10$^{-14}$, 1.24$\times$10$^{-6}$, and 4.07$\times$10$^{-5}$ for planets b, c, d, e, and f, respectively. These signal significance are still well below 0.1\% FAP albeit with less power considering two seasons of data are subtracted. The failure to falsify the planetary nature of the 26.9-day period combined with the fact we do not see the lunar cycle in any other HARPS activity indicators or seasons suggest such a signal is likely genuine. However, future high precision high cadence RV observations are recommended to support or refute this claim.

\subsubsection{Planet d's True Periodicity}

M11 reported planet d to have a period of 459.3 days, which is nearly half of the orbital period we have recovered in our analysis. However, we could only see a weak power of such a 459-day signal in our periodogram (see panel (f) in Figure \ref{fig:rvsearch}) and the signal is completely gone after we fit for the 959-day signal. We attempt to recover the 459-day periodicity by running the periodogram search only on the epochs used by M11. However, similar result is obtained where only the peak corresponding to the 959-day signal is recovered with FAP~=~9.52$\times$10$^{-5}$ with a peak containing weak power at around 459 days. We further create a synthetic dataset with an injected 459-day signal with parameters from M11 using the HARPS1 timestamps and errors to check whether such a signal could be retrieved. The periodogram is able to successfully detect the injected signal, with a weaker power around 1000-day periodicity. Our analysis suggests that the 459-day and 959-day periodicities could indeed be aliases of each other, depending on which one is the true signal. Considering the 459-day signal is recoverable if it is really present but none of our periodogram searches returned the 459-day signal, we therefore conclude the 959-day periodicity is the true period of this planet and what M11 recovered was an alias of the 959-day signal. Indeed, a model run with epoch and orbital parameters from M11 cannot reach MCMC chain convergence. Therefore, results derived from M11 are likely due to incomplete posterior sampling of orbital parameters.

\subsubsection{Aliasing or Independent?}

Orbital parameters we derived for the HD~134606 system show orbital periods of planets b, c, e, and f are close to integer ratios of each other. Such orbital configuration often times are targets of dynamical studies for their possibility of being in a resonance chain. Before such analysis can be conducted, one must rule out the case where they could be aliases of each other. We proceed by creating five new datasets, each dataset having one of the planetary signals subtracted from the original HARPS1 dataset using parameters from Table \ref{tab:param}. These datasets are then fed into a \texttt{RVSearch} to see if signals that are not subtracted can still be recovered. Indeed, all planets other than the one subtracted for the particular dataset are recovered with FAP well below 0.1\% FAP and we can safely assume the five signals in Table \ref{tab:param} are not aliases of each other.

\subsubsection{Cause of the RV Linear Trend}

\begin{figure*}[htbp!]
  \includegraphics[trim=0 0 0 0,clip,width=\textwidth]{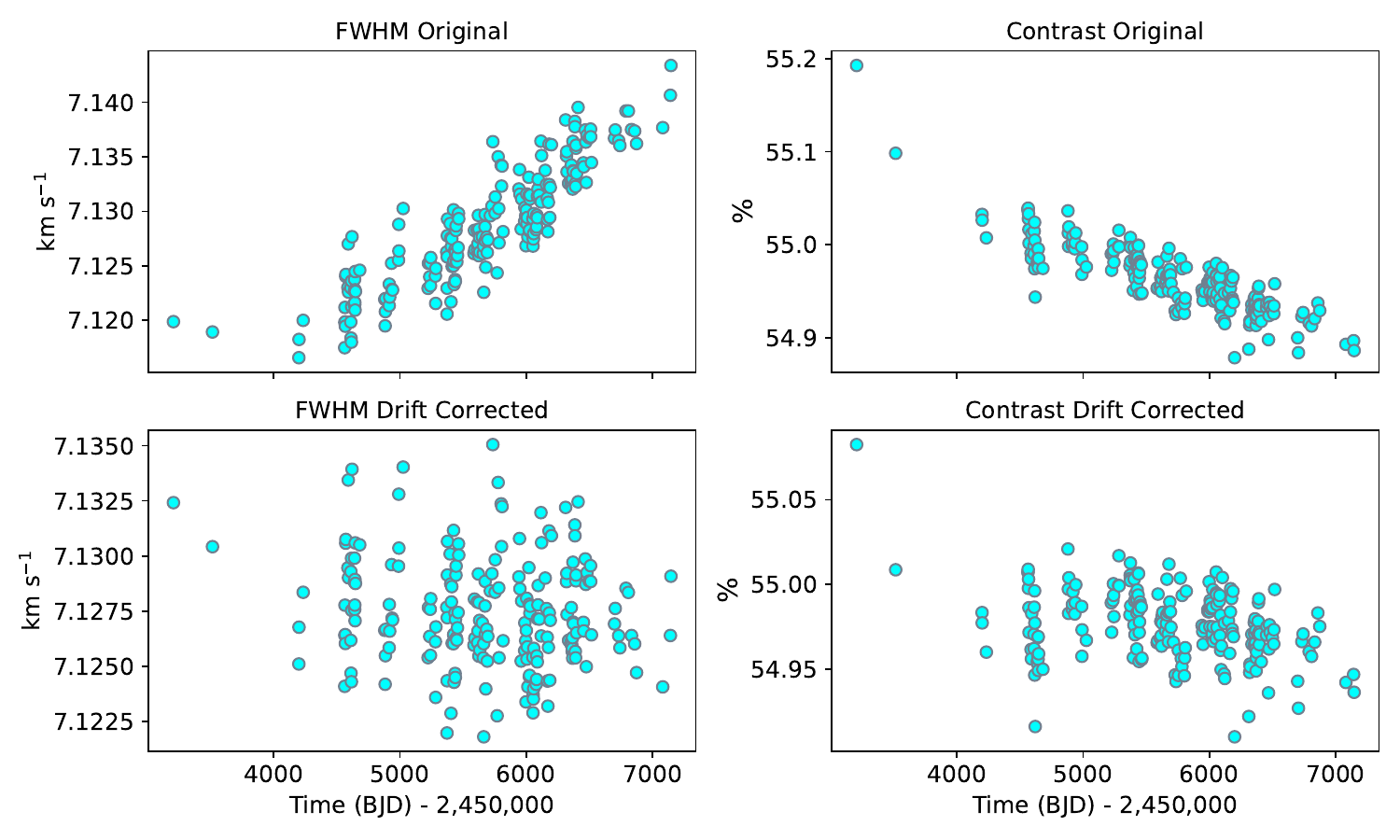}
  \caption{HARPS CCF FWHM and contrast data. Upper panels show the original time series of the two activity indicators and the lower panels show data that have been corrected for the systematic drift due to the instrument de-focusing issue. The linear trends seen in the original FWHM and contrast time series are completely instrumental.}
  \label{fig:drift}
\end{figure*}

As mentioned in section \ref{sec:rvsol}, a highly significant (17$\sigma$) linear trend is observed in the RV timeseries, which could be indicative of a very long period high mass companion. However, such a linear trend should be scrutinized when such a similar trend is also observed in activity indicators, which here is the case for HARPS dLW, CCF contrast, and FWHM. Both dLW and FWHM exhibit a positive trend while contrast shows a negative trend (see upper panels of Figure \ref{fig:drift}). The continued increase in line width and decrease in contrast point to an apparent long term drift in the line shape. One possible explanation is that a long-term magnetic cycle is driving such a change in the line shape. However, at the same time we do not see any change in the line skewness in the BIS indicator nor do we see a corresponding linear trend in the S-index, which usually is known to correlate well with magnetic cycles in G-type stars. Interestingly, numerous previous publications have observed similar long term drift in the HARPS CCF FWHM and contrast \citep{locurto2015,benatti2017,dumusque2018,barbato2019,costes2021,li2022} and such a drift has been attributed to the HARPS instrumental long-term de-focusing issue which was improved during the 2015 fibre upgrade. To remove the effect of the systematic drift, we apply two different polynomials below in Equations \ref{eqn:fwhm} and \ref{eqn:contrast} to FWHM and contrast, respectively, based on relations derived in \citet{costes2021} for G-type stars. The result can be seen in the bottom panels of Figure \ref{fig:drift} where the linear trends in both FWHM and contrast are completely removed after applying the drift correction. No more trend features persist in the FWHM and contrast data and we conclude that the linear trend we observe in the RV timeseries is indeed due to a physical long orbital period companion rather than a magnetic cycle.

\begin{multline}
\label{eqn:fwhm}
    \mathrm{FWHM_{corrected}} = \mathrm{FWHM_{original}} \\
    - (9.4\times10^{-10}\times (t-2452937.55)^{2} \\
    + 2.6\times10^{-6}\times (t-2452937.55) - 1.33\times10^{-2})
\end{multline}

\begin{multline}
\label{eqn:contrast}
    \mathrm{contrast_{corrected}} = \mathrm{contrast_{original}} \\
    - (7.0\times10^{-9}\times (t-2452937.55)^{2} \\
    - 7.2\times10^{-5}\times (t-2452937.55) + 1.29\times10^{-1})
\end{multline}

As suggested by our joint astrometry and RV analysis in section \ref{sec:rv+ast}, the companion is likely to be a high mass giant planet, brown dwarf, or a very low mass star. Our speckle imaging result in section \ref{sec:imagingsol} yields non-detection in the vicinity of the host star, ruling out presence of bright stellar companions within $\sim$1.2\arcsec{} of the star, or a little over 30~au of projected separation. The Washington Double Star Catalog however lists HD~134606 having a stellar companion of spectral type M3V at a projected angular separation of $\sim$57\arcsec{} that shares similar proper motion as the host star we are interested in. Assuming 57\arcsec{} being the minimum angular separation between HD~134606 and the M dwarf companion at a distance of 26.8~pc from Table \ref{tab:star}, we estimate the minimum orbital distance between the pair to be $\sim$1500~au. At this distance, the induced RV line-of-sight acceleration due to the M dwarf with a mass of 0.37~$M_{\sun}$ for a M3V star \citep{pecaut2013} using equation \ref{eqn:acc} can be estimated to be $\sim$0.03 m~s$^{-1}$~yr$^{-1}$, or $\sim$8$\times$10$^{-5}$ m~s$^{-1}$~d$^{-1}$, which is far smaller than the RV linear trend acceleration of 3.34$\times$10$^{-3}$ m~s$^{-1}$~d$^{-1}$ that we derived from Table \ref{tab:param}. Consequently, the distant M dwarf companion itself is not capable of producing the observed RV linear trend, and it is extremely likely that an additional sub-stellar companion is present in the outer regime of the HD~134606 system. Continued RV observations of this target are therefore needed over long term to resolve the orbit of this newly identified long period companion.

\subsection{Dynamical Analysis}
\label{sec:dyna}

\begin{figure}[htbp!]
  \includegraphics[trim=0 0 0 0,clip,width=\columnwidth]{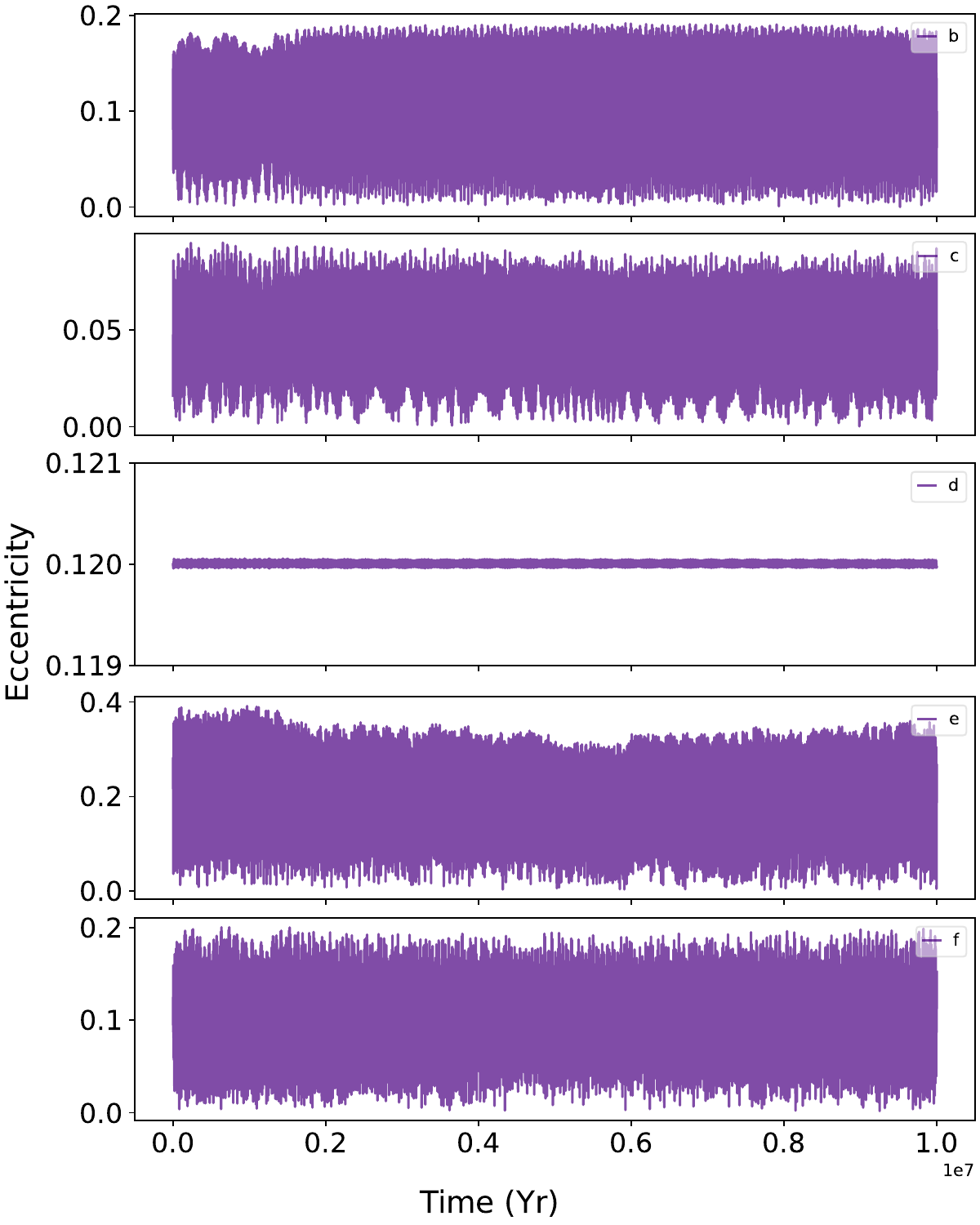}
  \caption{Dynamical simulation result of the HD~134606 system with five planets. Eccentricities for each planet were recorded for a simulation duration of 10 million years. The inner planets experience significant eccentricity variations.}
  \label{fig:dynaecc}
\end{figure}

The inner four planets b, c, e, and f orbit closer to each other in the inner regime of the system and have orbital periods near integer ratios. To explore the dynamics of the system, we conduct an N-body simulation using the \texttt{REBOUND} package \citep{rein2012a}. The system is initialized using our derived stellar mass value from Table \ref{tab:star} and the maximum likelihood orbital parameters for the planets from Table \ref{tab:param} with planetary orbits assumed to be coplanar. The system is integrated using the symplectic integrator \texttt{WHFast} \citep{rein2015c} with a time step of 1/20 of the inner most planet's orbital period, or 4.8 hours, consistent with the recommended step size from \citet{duncan1998} to ensure proper time resolution orbit sampling. The integration duration is set to 10 million years and we record the eccentricity values for all the planets every 100 years. The system is dynamically stable over the duration of the simulation as shown by the eccentricity variation of planets in Figure \ref{fig:dynaecc}. However, the inner four planets appear to be experiencing high levels of mutual interaction as evident by the mild eccentricity variation of planet c from near circular to up to 0.1 eccentricity. Planet b, e, and f exhibit higher amounts of eccentricity variation, with planet e showcasing variation up to 0.4. Given the low mass of planet e, its proximity to the G-type star, and the age of the system, the observed eccentricity variation is unlikely to have inherited from the formation of the planetary system since tidal circularization would have already dampened the orbital eccentricity of planet e. Therefore, the observed moderate eccentricity of planet e opens up a possible scenario where the planet's orbit may have been perturbed by a dynamical event either internal or external to the system. If planet e is indeed present, the 0.4 eccentricity variation could be the key to the intriguing dynamical history of this system. Based on our dynamical result, planet e exhibits orbital eccentricity of currently observed value of 0.2 or smaller for 58\% of the sampled time steps, suggesting observation of the current eccentricity or smaller is not a low probability event. However, more high cadence follow-up observations are needed to further validate the presence of this planet and refine its orbital eccentricity (presently with 0.14 uncertainty) such that the current system configuration can be used for future dynamical studies with less ambiguity.

The orbital periods of the inner four planets are near-integer ratios of each other. Specifically, planet e ($\sim$4 days) and b ($\sim$12 days)'s periods are near ratios of 1:3, and planet b ($\sim$12 days), f ($\sim$27 days), and c ($\sim$59 days)'s periods are close to a 1:2:4 ratio. To study the dynamical interaction between these pairs, we conduct a separate N-body simulation for a duration of 1,000 years and record evolution of resonant angles for these pairs every 0.1 year. Unsurprisingly, given the amount of deviation from the perfect period integer ratios, none of the resonant angles exhibit evidence of libration and the four inner planets therefore are not in orbital resonance with each other, consistent with previous results that planets do not show much preference to be near mean-motion resonances \citep{fabrycky2014}.

\subsection{Direct Imaging Prospects}
\label{sec:DIprosp}

\begin{figure*}[htbp!]
    \begin{center}
        \begin{tabular}{cc}
            \includegraphics[trim=40 7 20 20,clip,width=0.5\textwidth]{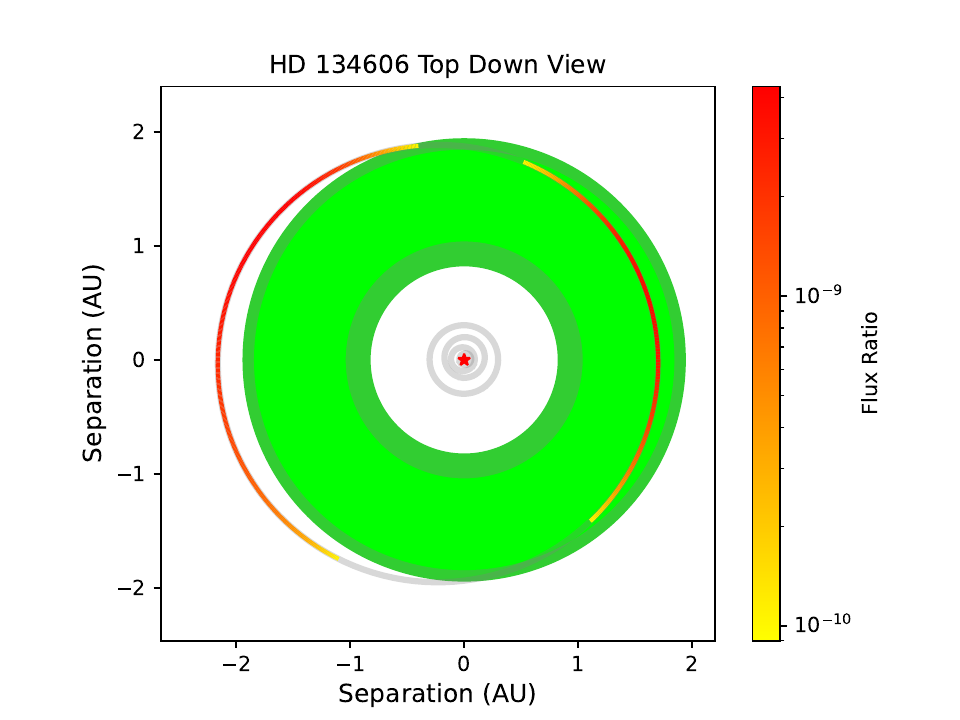} &
            \includegraphics[trim=40 7 20 20,clip,width=0.5\textwidth]{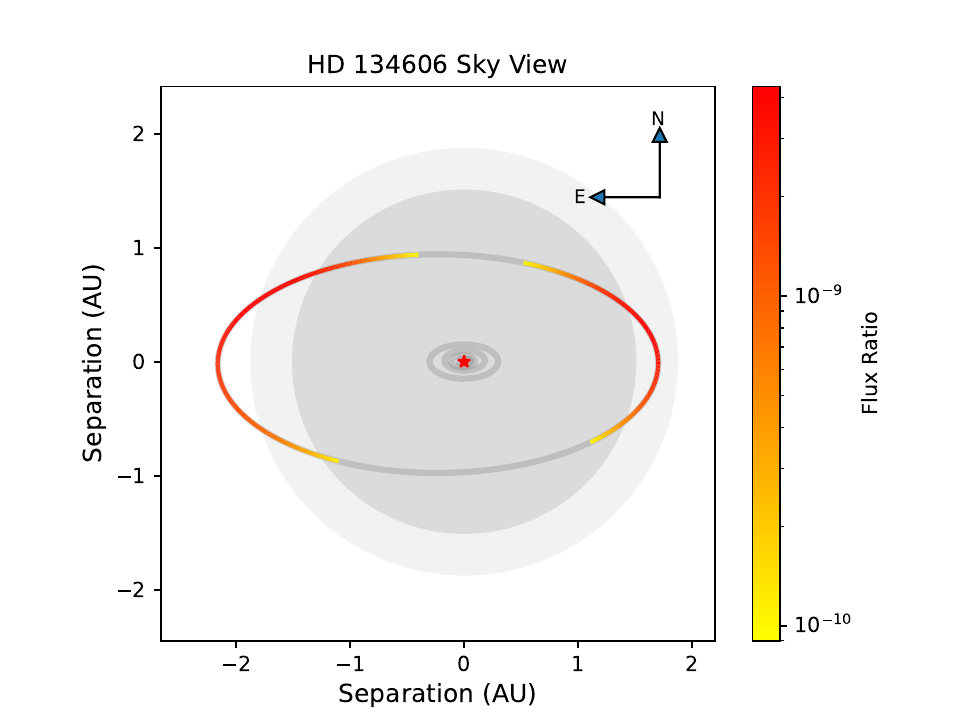} \\
        \end{tabular}
    \end{center}
    \caption{Direct imaging visibility of the five planets in the HD~134606 system. Left panel: top down view of the system with CHZ (bright green) and OHZ (dark green) overlaid. Right panel: sky projected view of the system with 60$^{\circ}$ orbital inclination. IWA and IWA$_{0.5}$ are represented by the bright and dark shaded disks, respectively. Orbits are color coded if they are above the 1$\times$10$^{-10}$ flux ratio with 60$^{\circ}$ inclination. Inner four planetary orbits are too compact to be shown with detail on this scale.}
    \label{fig:planetfflux}
\end{figure*}

Future space-based direct imaging missions such as the Habitable Worlds Observatory will be targeting terrestrial exoplanets within the habitable zone (HZ) of nearby stars \citep{kasting1993a,kane2012a,kopparapu2013a,kopparapu2014,kane2016c,hill2018,hill2023}, opening doors to direct atmospheric retrieval of low mass planets as well as direct detection of such planets that are hard or impossible to be discovered using other detection methods. Here, we estimate the feasibility of directly imaging planets in the HD~134606 system using a first-order flux ratio estimation with telescope configuration of the Habitable Exoplanet Observatory (HabEx) as a proxy to the next generation Habitable Worlds Observatory as recommended by the Decadal Survey of Astronomy and Astrophysics 2020 \citep{nas2021}. We follow the methodology presented in \citet{li2021} that used the required 1$\times$10$^{-10}$ contrast as the detection limit for the HabEx mission with starshade \citep{gaudi2020} and taking into consideration the transmittance profile of starshade for the flux ratio estimation. Detectability of planets in an imaging mission also depends on their angular separations. The inner working angle (IWA) of HabEx, which is the angular size of the starshade as seen from the telescope, is taken to be 70 mas. In addition, we employ a relatively optimistic angular separation limit, IWA$_{0.5}$, which is defined as the angular radius from the starshade center where the transmittance is 50\%. For HabEx, IWA$_{0.5}$ is estimated to be at 56.4 mas when averaged across all wavelengths and can serve as the minimum angular separation where exoplanet detections can be made \citep{gaudi2020, li2021}. Our flux ratio estimation assumes Lambertian reflectance models and selects relatively conservative geometric albedo values of 0.2 in the visible band for all the planets. Planetary radius values are estimated using mass-radius relationship from \citet{chen2017}. Using stellar and orbital parameters from Tables~\ref{tab:star} and \ref{tab:param} along with an assumed inclination value of 60$^{\circ}$, we present the estimation in Figure \ref{fig:planetfflux}.

The planetary masses and radii are adjusted according to the inclination value and the system is confirmed to be stable at this orbital inclination. Clearly, the inner four planets are too close to the star and are completely blocked by the starshade as indicated by the two shaded disks representing IWA and IWA$_{0.5}$ in the right panel. However, the outermost planet d appears to be bright enough and with its orbit wide enough to be detected by imaging, especially near maximum angular separation. Maximum brightness of this planet is achieved around $\sim$4.7$\times$10$^{-9}$ planet-to-star flux ratio at locations when the planet is near the edge of the IWA (east of the star), making it a promising candidate for future direct imaging missions. The conclusion of direct imaging feasibility of planet d still stands if we assign even lower geometric albedo values and more edge-on inclination angles. Interestingly, as indicated by the left panel in Figure \ref{fig:planetfflux}, over half of the planet's orbit is in the optimistic HZ (OHZ) and a significant portion of it in the conservative HZ (CHZ), the boundaries of which are defined in \citet{kopparapu2013a,kopparapu2014}. Although planet d is unlikely to be habitable considering its relatively large mass and size, it could still serve as an interesting target for future studies of habitability of exomoons around giant planets in the HZ.

\begin{figure}[htbp!]
  \includegraphics[trim=5 5 5 5,clip,width=\columnwidth]{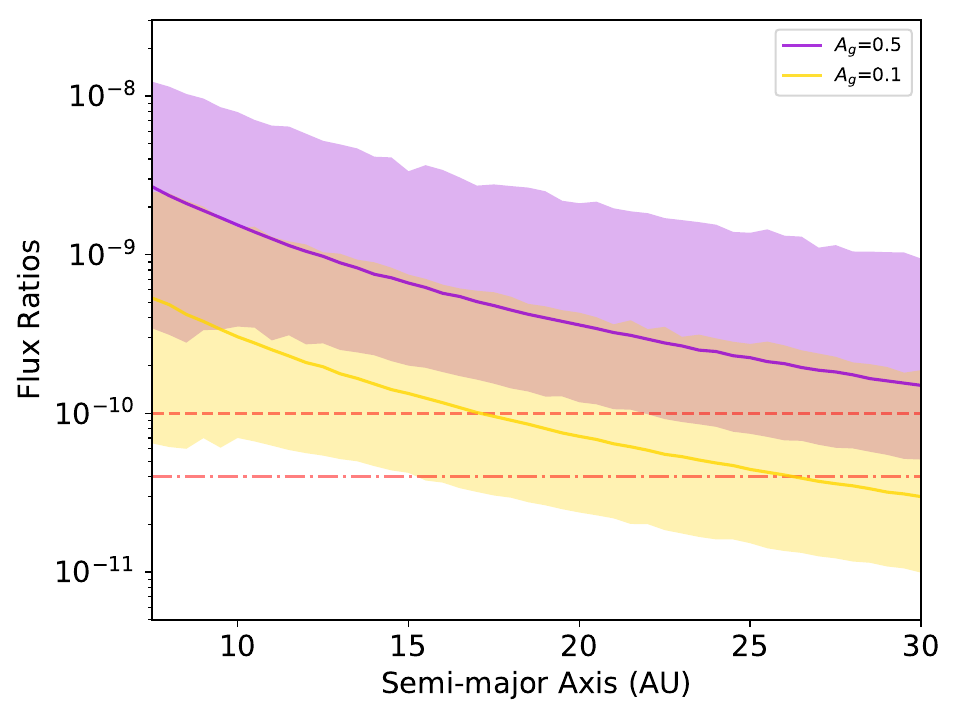}
  \caption{Flux ratio estimate as a function of semi-major axis of the long period companion based on random sampling of orbital parameters. The horizontal dashed line and the dot-dashed line represent the required and optimistic contrast ratio limit of future space-based direct imaging missions.}
  \label{fig:fluxratio}
\end{figure}

As mentioned previously in section \ref{sec:nature}, the origin of the observed RV linear trend is due to a physical companion in a very long period orbit. According to Figure \ref{fig:limits} from the combined RV and astrometry analysis, such a companion could either be a very large mass gas giant planet, a brown dwarf, or a very low mass stellar companion, depending on the orbital separation of such an object. Based on the angular separation and derived mass information from the RV and astrometry analysis, we estimate the imaging feasibility of such a companion as a function of orbital semimajor axis by creating a pool of synthetic samples. We provide such an estimate between 7.5~au and 30~au with a step of 0.5~au. At each semimajor axis step, we randomly draw 10,000 exoplanet samples with eccentricity values drawn from a beta distribution \citep{kipping2013b}, orbital inclination from a uniform distribution of cos$i$, and with argument of periastron, longitude of periastron, and true anomaly all drawn from uniform distributions. For each drawn sample, on-sky angular projection is calculated and this value is passed into an interpolation function based on the median companion mass values as a function of angular separation from the RV and astrometry analysis in Figure \ref{fig:limits} to yield an estimated companion mass value at this semimajor axis. The derived mass is then passed into mass-radius relationship from \citet{chen2017} and the flux ratio calculation then followed the same procedure as above using the drawn orbital parameters with two different geometric albedo values $A_{g}$: 0.1, and 0.5, to take into consideration that the companion could either be a gas giant or a brown dwarf. These values are consistent with albedo models for solar system objects \citep{roberge2017} and brown dwarfs \citep{marley1999}. Since we are using a reflectance only model, we place an upper limit of 80 Jupiter mass on the mass of the samples drawn. The above process is repeated for all 10,000 samples at each semimajor axis location. If for certain samples the derived mass exceeded the upper limit, we discard such samples and redraw from the distributions such that all 10,000 samples are below the stellar mass limit. Finally, we obtain a distribution of 10,000 flux ratio values for each of the orbital locations from 7.5 to 30~au. At each semimajor axis location, we take 95\% confidence interval along with the median value from the estimated flux ratio distribution as the brightness range of the companion as a function of different orbit sizes. The result is shown in Figure \ref{fig:fluxratio}. The horizontal dashed and dot-dashed lines represent the required 1$\times$10$^{-10}$ and optimistic 4$\times$10$^{-11}$ contrast limit of HabEx. At an orbital distance of $\sim$30~au, $\sim$50\% of the samples could potentially be directly imaged assuming the required contrast limit and $A_g$ of 0.5. For $A_g$ of 0.1, the sample portion drops to $<$~20\%. As expected, planet-to-star reflected light flux ratio of this companion greatly diminishes the farther away the possible location of the companion is from the star and the detectability of such a companion through direct imaging in the visible band is low unless it orbits closer to the star. However, as the expected companion mass increases as a function of separation (see Fig. \ref{fig:limits}), the companion would be massive enough at large separations (beyond 20-30~au) at which thermal emission would dominate and therefore be visible in the infrared wavelength. At even larger distances (beyond 60-70~au), the companion nature is likely to be a low mass star and would be self-luminous.

\subsection{Planet Naming Convention}
\label{sec:naming}

According to the policy adopted by NASA Exoplanet Archive regarding the inclusion of exoplanets in the archive, results of a study must be peer-reviewed and accepted for publication in literature. The HD~134606 system was originally reported to host three planets by M11. Although the M11 paper was not refereed and only a table of orbital parameters without further analysis was given, we respect the b, c, and d letter assignment to the 12-day, 59-day, and 959-day (originally 459 days by M11) planets. Because of this, despite the 4-day and 27-day planets having shorter orbital periods, we assign them with letters e, and f, respectively, instead of reassigning designations for all planets according to their orbital distances from the star. This designation is consistent to the ordering presented in Table \ref{tab:param}.


\section{Conclusions}
\label{sec:conclusions}

In this work, we revisited one of the multi-planet systems observed by HARPS. Assisted by the continued HARPS observations since the M11 paper, the addition of UCLES data, and the re-reduction of the HARPS data by \citet{trifonov2020}, we are able to significantly revise the architecture of this system with updated orbital parameters and newly discovered planets. In total, 285 RV observations spanning a total of over 19 years were utilized in the RV modeling process, resulting in a total of five planets detected in the RV time series with great statistical significance. Out of the five detected planets, we confirmed planets b and c from M11 and revised planet d's periodicity to 959 days and deemed the result from M11 was likely an aliasing issue due to insufficient posterior sampling of orbital parameters. Two new super-Earths e and f were discovered with periods of 4 and 27 days, respectively. Additionally, we identified a significant linear trend in the RV residual and confirmed its origin most likely to be a new sub-stellar companion. Major steps were undertaken in an attempt to falsify the planetary nature of the recovered signals through analysis of various stellar activity indicators and RV sampling. No stellar rotation signatures can be recovered from spectroscopy or photometry and the star was deemed quiet. None of the suspicious periodicities in the activity time series would rule out the likelihood of the five RV signals being planets.

Multiple detection methods were employed in this study. Although both photometry and imaging yielded negative results, photometry helped confirm the quiet nature of the star whereas imaging placed upper limit constraints on the mass of the newly identified companion in the outer regime of the planetary system. Using combined RV and astrometry analysis, we were able to derive mass estimates as a function of angular separation, which later was used to estimate the feasibility of directly imaging this distant companion causing the RV linear trend and astrometric acceleration. The HD~134606 multi-planet system is now characterized to host 5 planets, with the inner four being low mass super-Earths and mini-Neptunes orbiting in a relatively compact inner region, with another gas giant in the HZ. Further out, a sub-stellar companion resides in a very wide orbit unresolved by RVs and astrometry and an additional M-dwarf star at a sufficiently large distance that we see no evidence of it in the RVs and astrometry. The different characteristics of orbiting bodies within the system prompts interesting questions regarding the possible formation scenarios and the pathways it took to reach the current observed architecture. The eccentricity variation of the inner four planets may provide some clues on the evolution and recent dynamical history of the system. Dedicated long-term high-precision RV monitoring of HD~134606 is required to further verify the inner planets and resolve the orbital characteristics of the sub-stellar companion to hopefully get a glimpse of the true nature of this object. Direct imaging of this companion is certainly possible with either reflected light or thermal emission from this companion. Planet d in addition offers itself as an exciting target for future imaging missions thanks to its relatively large size, favorable planet-to-star flux ratio, and orbital location being in the habitable zone. HD~134606 provides plenty of excitement and opportunities for future studies and follow up observations.


\section*{Acknowledgements}

The authors would like to thank the anonymous referee for the valuable feedback. Z. L. wishes to thank Mirek G. Brandt for the many ideas regarding the astrometric analysis of this target, Howard Isaacson for the discussion on RVs and stellar activities, and Alex Venner for the conversation on HARPS RVs and systematics. P. D. acknowledges support by a 51 Pegasi b Postdoctoral Fellowship from the Heising-Simons Foundation. This work is based in part on data acquired at the Anglo-Australian Telescope. We acknowledge the traditional custodians of the land on which the AAT stands, the Gamilaraay people, and pay our respects to elders past and present. Dynamical simulations in this paper made use of the \texttt{REBOUND} code which is freely available at \url{http://github.com/hannorein/rebound}. This research has made use of the NASA Exoplanet Archive, which is operated by the California Institute of Technology, under contract with the National Aeronautics and Space Administration under the Exoplanet Exploration Program. This research has also made use of the Washington Double Star Catalog maintained at the U.S. Naval Observatory.


\software{\texttt{ACTIN 2} \citep{gsilva2018,gsilva2021}, \texttt{EXOFASTv2} \citep{eastman2013, eastman2019}, \texttt{GLS} \citep{zechmeister2009}, \texttt{RadVel} \citep{fulton2018a}, \texttt{REBOUND} \citep{rein2012a}, \texttt{RVSearch} \citep{rosenthal2021}}, \texttt{SpecMath-Emp} \citep{yee2017}



\begin{thebibliography}{}
\expandafter\ifx\csname natexlab\endcsname\relax\def\natexlab#1{#1}\fi
\providecommand{\url}[1]{\href{#1}{#1}}
\providecommand{\dodoi}[1]{doi:~\href{http://doi.org/#1}{\nolinkurl{#1}}}
\providecommand{\doeprint}[1]{\href{http://ascl.net/#1}{\nolinkurl{http://ascl.net/#1}}}
\providecommand{\doarXiv}[1]{\href{https://arxiv.org/abs/#1}{\nolinkurl{https://arxiv.org/abs/#1}}}

\bibitem[{esa(1997)}]{esa1997}
 1997, ESA Special Publication, Vol. 1200, {The HIPPARCOS and TYCHO catalogues.
  Astrometric and photometric star catalogues derived from the ESA HIPPARCOS
  Space Astrometry Mission}

\bibitem[{{Barbato} {et~al.}(2019){Barbato}, {Sozzetti}, {Biazzo}, {Malavolta},
  {Santos}, {Damasso}, {Lanza}, {Pinamonti}, {Affer}, {Benatti}, {Bignamini},
  {Bonomo}, {Borsa}, {Carleo}, {Claudi}, {Cosentino}, {Covino}, {Desidera},
  {Esposito}, {Giacobbe}, {Gonz{\'a}lez-{\'A}lvarez}, {Gratton}, {Harutyunyan},
  {Leto}, {Maggio}, {Maldonado}, {Mancini}, {Masiero}, {Micela}, {Molinari},
  {Nascimbeni}, {Pagano}, {Piotto}, {Poretti}, {Rainer}, {Scandariato},
  {Smareglia}, {Colombo}, {Di Fabrizio}, {Faria}, {Martinez Fiorenzano},
  {Molinaro}, \& {Pedani}}]{barbato2019}
{Barbato}, D., {Sozzetti}, A., {Biazzo}, K., {et~al.} 2019, \aap, 621, A110,
  \dodoi{10.1051/0004-6361/201834305}

\bibitem[{{Benatti} {et~al.}(2017){Benatti}, {Desidera}, {Damasso},
  {Malavolta}, {Lanza}, {Biazzo}, {Bonomo}, {Claudi}, {Marzari}, {Poretti},
  {Gratton}, {Micela}, {Pagano}, {Piotto}, {Sozzetti}, {Boccato}, {Cosentino},
  {Covino}, {Maggio}, {Molinari}, {Smareglia}, {Affer}, {Andreuzzi},
  {Bignamini}, {Borsa}, {di Fabrizio}, {Esposito}, {Martinez Fiorenzano},
  {Messina}, {Giacobbe}, {Harutyunyan}, {Knapic}, {Maldonado}, {Masiero},
  {Nascimbeni}, {Pedani}, {Rainer}, {Scandariato}, \& {Silvotti}}]{benatti2017}
{Benatti}, S., {Desidera}, S., {Damasso}, M., {et~al.} 2017, \aap, 599, A90,
  \dodoi{10.1051/0004-6361/201629484}

\bibitem[{{Brandt}(2018)}]{brandt2018}
{Brandt}, T.~D. 2018, \apjs, 239, 31, \dodoi{10.3847/1538-4365/aaec06}

\bibitem[{{Brandt}(2021)}]{brandt2021}
---. 2021, \apjs, 254, 42, \dodoi{10.3847/1538-4365/abf93c}

\bibitem[{{Butler} {et~al.}(1996){Butler}, {Marcy}, {Williams}, {McCarthy},
  {Dosanjh}, \& {Vogt}}]{butler1996b}
{Butler}, R.~P., {Marcy}, G.~W., {Williams}, E., {et~al.} 1996, \pasp, 108,
  500, \dodoi{10.1086/133755}

\bibitem[{{Chambers} {et~al.}(1996){Chambers}, {Wetherill}, \&
  {Boss}}]{chambers1996}
{Chambers}, J.~E., {Wetherill}, G.~W., \& {Boss}, A.~P. 1996, \icarus, 119,
  261, \dodoi{10.1006/icar.1996.0019}

\bibitem[{{Chen} \& {Kipping}(2017)}]{chen2017}
{Chen}, J., \& {Kipping}, D. 2017, \apj, 834, 17,
  \dodoi{10.3847/1538-4357/834/1/17}

\bibitem[{{Choi} {et~al.}(2016){Choi}, {Dotter}, {Conroy}, {Cantiello},
  {Paxton}, \& {Johnson}}]{choi2016}
{Choi}, J., {Dotter}, A., {Conroy}, C., {et~al.} 2016, \apj, 823, 102,
  \dodoi{10.3847/0004-637X/823/2/102}

\bibitem[{{Clanton} \& {Gaudi}(2016)}]{clanton2016}
{Clanton}, C., \& {Gaudi}, B.~S. 2016, \apj, 819, 125,
  \dodoi{10.3847/0004-637X/819/2/125}

\bibitem[{{Costes} {et~al.}(2021){Costes}, {Watson}, {de Mooij}, {Saar},
  {Dumusque}, {Cameron}, {Phillips}, {G{\"u}nther}, {Jenkins}, {Mortier}, \&
  {Thompson}}]{costes2021}
{Costes}, J.~C., {Watson}, C.~A., {de Mooij}, E., {et~al.} 2021, \mnras, 505,
  830, \dodoi{10.1093/mnras/stab1183}

\bibitem[{{Dalba} {et~al.}(2021){Dalba}, {Kane}, {Howell}, {Horch}, {Li},
  {Hirsch}, {Burt}, {Brandt}, {Mo{\v{c}}nik}, {Henry}, {Everett}, {Rosenthal},
  \& {Howard}}]{dalba2021}
{Dalba}, P.~A., {Kane}, S.~R., {Howell}, S.~B., {et~al.} 2021, \aj, 161, 123,
  \dodoi{10.3847/1538-3881/abd6ed}

\bibitem[{{Dawson} \& {Fabrycky}(2010)}]{dawson2010}
{Dawson}, R.~I., \& {Fabrycky}, D.~C. 2010, \apj, 722, 937,
  \dodoi{10.1088/0004-637X/722/1/937}

\bibitem[{{Diego} {et~al.}(1990){Diego}, {Charalambous}, {Fish}, \&
  {Walker}}]{diego1990}
{Diego}, F., {Charalambous}, A., {Fish}, A.~C., \& {Walker}, D.~D. 1990, in
  \procspie, Vol. 1235, Instrumentation in Astronomy VII, ed. D.~L. {Crawford},
  562--576, \dodoi{10.1117/12.19119}

\bibitem[{{Dotter}(2016)}]{dotter2016}
{Dotter}, A. 2016, \apjs, 222, 8, \dodoi{10.3847/0067-0049/222/1/8}

\bibitem[{{Dumusque}(2018)}]{dumusque2018}
{Dumusque}, X. 2018, \aap, 620, A47, \dodoi{10.1051/0004-6361/201833795}

\bibitem[{{Dumusque} {et~al.}(2011){Dumusque}, {Udry}, {Lovis}, {Santos}, \&
  {Monteiro}}]{dumusque2011}
{Dumusque}, X., {Udry}, S., {Lovis}, C., {Santos}, N.~C., \& {Monteiro},
  M.~J.~P.~F.~G. 2011, \aap, 525, A140, \dodoi{10.1051/0004-6361/201014097}

\bibitem[{{Duncan} {et~al.}(1991){Duncan}, {Vaughan}, {Wilson}, {Preston},
  {Frazer}, {Lanning}, {Misch}, {Mueller}, {Soyumer}, {Woodard}, {Baliunas},
  {Noyes}, {Hartmann}, {Porter}, {Zwaan}, {Middelkoop}, {Rutten}, \&
  {Mihalas}}]{duncan1991}
{Duncan}, D.~K., {Vaughan}, A.~H., {Wilson}, O.~C., {et~al.} 1991, \apjs, 76,
  383, \dodoi{10.1086/191572}

\bibitem[{{Duncan} {et~al.}(1998){Duncan}, {Levison}, \& {Lee}}]{duncan1998}
{Duncan}, M.~J., {Levison}, H.~F., \& {Lee}, M.~H. 1998, \aj, 116, 2067,
  \dodoi{10.1086/300541}

\bibitem[{{Eastman} {et~al.}(2013){Eastman}, {Gaudi}, \& {Agol}}]{eastman2013}
{Eastman}, J., {Gaudi}, B.~S., \& {Agol}, E. 2013, \pasp, 125, 83,
  \dodoi{10.1086/669497}

\bibitem[{{Eastman} {et~al.}(2019){Eastman}, {Rodriguez}, {Agol}, {Stassun},
  {Beatty}, {Vanderburg}, {Gaudi}, {Collins}, \& {Luger}}]{eastman2019}
{Eastman}, J.~D., {Rodriguez}, J.~E., {Agol}, E., {et~al.} 2019, arXiv
  e-prints, arXiv:1907.09480.
\newblock \doarXiv{1907.09480}

\bibitem[{{Fabrycky} {et~al.}(2014){Fabrycky}, {Lissauer}, {Ragozzine}, {Rowe},
  {Steffen}, {Agol}, {Barclay}, {Batalha}, {Borucki}, {Ciardi}, {Ford},
  {Gautier}, {Geary}, {Holman}, {Jenkins}, {Li}, {Morehead}, {Morris},
  {Shporer}, {Smith}, {Still}, \& {Van Cleve}}]{fabrycky2014}
{Fabrycky}, D.~C., {Lissauer}, J.~J., {Ragozzine}, D., {et~al.} 2014, \apj,
  790, 146, \dodoi{10.1088/0004-637X/790/2/146}

\bibitem[{{Fetherolf} {et~al.}(2023){Fetherolf}, {Pepper}, {Simpson}, {Kane},
  {Mo{\v{c}}nik}, {English}, {Antoci}, {Huber}, {Jenkins}, {Stassun},
  {Twicken}, {Vanderspek}, \& {Winn}}]{fetherolf2023}
{Fetherolf}, T., {Pepper}, J., {Simpson}, E., {et~al.} 2023, \apjs, 268, 4,
  \dodoi{10.3847/1538-4365/acdee5}

\bibitem[{{Fischer} {et~al.}(2016){Fischer}, {Anglada-Escude}, {Arriagada},
  {Baluev}, {Bean}, {Bouchy}, {Buchhave}, {Carroll}, {Chakraborty}, {Crepp},
  {Dawson}, {Diddams}, {Dumusque}, {Eastman}, {Endl}, {Figueira}, {Ford},
  {Foreman-Mackey}, {Fournier}, {F{\H{u}}r{\'e}sz}, {Gaudi}, {Gregory},
  {Grundahl}, {Hatzes}, {H{\'e}brard}, {Herrero}, {Hogg}, {Howard}, {Johnson},
  {Jorden}, {Jurgenson}, {Latham}, {Laughlin}, {Loredo}, {Lovis}, {Mahadevan},
  {McCracken}, {Pepe}, {Perez}, {Phillips}, {Plavchan}, {Prato}, {Quirrenbach},
  {Reiners}, {Robertson}, {Santos}, {Sawyer}, {Segransan}, {Sozzetti},
  {Steinmetz}, {Szentgyorgyi}, {Udry}, {Valenti}, {Wang}, {Wittenmyer}, \&
  {Wright}}]{fischer2016}
{Fischer}, D.~A., {Anglada-Escude}, G., {Arriagada}, P., {et~al.} 2016, \pasp,
  128, 066001, \dodoi{10.1088/1538-3873/128/964/066001}

\bibitem[{{Ford}(2014)}]{ford2014}
{Ford}, E.~B. 2014, Proceedings of the National Academy of Science, 111, 12616,
  \dodoi{10.1073/pnas.1304219111}

\bibitem[{{Fulton} {et~al.}(2018){Fulton}, {Petigura}, {Blunt}, \&
  {Sinukoff}}]{fulton2018a}
{Fulton}, B.~J., {Petigura}, E.~A., {Blunt}, S., \& {Sinukoff}, E. 2018, \pasp,
  130, 044504, \dodoi{10.1088/1538-3873/aaaaa8}

\bibitem[{{Fulton} {et~al.}(2021){Fulton}, {Rosenthal}, {Hirsch}, {Isaacson},
  {Howard}, {Dedrick}, {Sherstyuk}, {Blunt}, {Petigura}, {Knutson}, {Behmard},
  {Chontos}, {Crepp}, {Crossfield}, {Dalba}, {Fischer}, {Henry}, {Kane},
  {Kosiarek}, {Marcy}, {Rubenzahl}, {Weiss}, \& {Wright}}]{fulton2021}
{Fulton}, B.~J., {Rosenthal}, L.~J., {Hirsch}, L.~A., {et~al.} 2021, \apjs,
  255, 14, \dodoi{10.3847/1538-4365/abfcc1}

\bibitem[{{Gaia Collaboration} {et~al.}(2018){Gaia Collaboration}, {Brown},
  {Vallenari}, {Prusti}, {de Bruijne}, {Babusiaux}, {Bailer-Jones}, {Biermann},
  {Evans}, {Eyer}, {Jansen}, {Jordi}, {Klioner}, {Lammers}, {Lindegren},
  {Luri}, {Mignard}, {Panem}, {Pourbaix}, {Randich}, {Sartoretti}, {Siddiqui},
  {Soubiran}, {van Leeuwen}, {Walton}, {Arenou}, {Bastian}, {Cropper},
  {Drimmel}, {Katz}, {Lattanzi}, {Bakker}, {Cacciari}, {Casta{\~n}eda},
  {Chaoul}, {Cheek}, {De Angeli}, {Fabricius}, {Guerra}, {Holl}, {Masana},
  {Messineo}, {Mowlavi}, {Nienartowicz}, {Panuzzo}, {Portell}, {Riello},
  {Seabroke}, {Tanga}, {Th{\'e}venin}, {Gracia-Abril}, {Comoretto},
  {Garcia-Reinaldos}, {Teyssier}, {Altmann}, {Andrae}, {Audard},
  {Bellas-Velidis}, {Benson}, {Berthier}, {Blomme}, {Burgess}, {Busso},
  {Carry}, {Cellino}, {Clementini}, {Clotet}, {Creevey}, {Davidson}, {De
  Ridder}, {Delchambre}, {Dell'Oro}, {Ducourant},
  {Fern{\'a}ndez-Hern{\'a}ndez}, {Fouesneau}, {Fr{\'e}mat}, {Galluccio},
  {Garc{\'\i}a-Torres}, {Gonz{\'a}lez-N{\'u}{\~n}ez}, {Gonz{\'a}lez-Vidal},
  {Gosset}, {Guy}, {Halbwachs}, {Hambly}, {Harrison}, {Hern{\'a}ndez},
  {Hestroffer}, {Hodgkin}, {Hutton}, {Jasniewicz}, {Jean-Antoine-Piccolo},
  {Jordan}, {Korn}, {Krone-Martins}, {Lanzafame}, {Lebzelter}, {L{\"o}ffler},
  {Manteiga}, {Marrese}, {Mart{\'\i}n-Fleitas}, {Moitinho}, {Mora}, {Muinonen},
  {Osinde}, {Pancino}, {Pauwels}, {Petit}, {Recio-Blanco}, {Richards},
  {Rimoldini}, {Robin}, {Sarro}, {Siopis}, {Smith}, {Sozzetti}, {S{\"u}veges},
  {Torra}, {van Reeven}, {Abbas}, {Abreu Aramburu}, {Accart}, {Aerts},
  {Altavilla}, {{\'A}lvarez}, {Alvarez}, {Alves}, {Anderson}, {Andrei},
  {Anglada Varela}, {Antiche}, {Antoja}, {Arcay}, {Astraatmadja}, {Bach},
  {Baker}, {Balaguer-N{\'u}{\~n}ez}, {Balm}, {Barache}, {Barata}, {Barbato},
  {Barblan}, {Barklem}, {Barrado}, {Barros}, {Barstow}, {Bartholom{\'e}
  Mu{\~n}oz}, {Bassilana}, {Becciani}, {Bellazzini}, {Berihuete}, {Bertone},
  {Bianchi}, {Bienaym{\'e}}, {Blanco-Cuaresma}, {Boch}, {Boeche}, {Bombrun},
  {Borrachero}, {Bossini}, {Bouquillon}, {Bourda}, {Bragaglia}, {Bramante},
  {Breddels}, {Bressan}, {Brouillet}, {Br{\"u}semeister}, {Brugaletta},
  {Bucciarelli}, {Burlacu}, {Busonero}, {Butkevich}, {Buzzi}, {Caffau},
  {Cancelliere}, {Cannizzaro}, {Cantat-Gaudin}, {Carballo}, {Carlucci},
  {Carrasco}, {Casamiquela}, {Castellani}, {Castro-Ginard}, {Charlot},
  {Chemin}, {Chiavassa}, {Cocozza}, {Costigan}, {Cowell}, {Crifo}, {Crosta},
  {Crowley}, {Cuypers}, {Dafonte}, {Damerdji}, {Dapergolas}, {David}, {David},
  {de Laverny}, {De Luise}, {De March}, {de Martino}, {de Souza}, {de Torres},
  {Debosscher}, {del Pozo}, {Delbo}, {Delgado}, {Delgado}, {Di Matteo},
  {Diakite}, {Diener}, {Distefano}, {Dolding}, {Drazinos}, {Dur{\'a}n},
  {Edvardsson}, {Enke}, {Eriksson}, {Esquej}, {Eynard Bontemps}, {Fabre},
  {Fabrizio}, {Faigler}, {Falc{\~a}o}, {Farr{\`a}s Casas}, {Federici},
  {Fedorets}, {Fernique}, {Figueras}, {Filippi}, {Findeisen}, {Fonti},
  {Fraile}, {Fraser}, {Fr{\'e}zouls}, {Gai}, {Galleti}, {Garabato},
  {Garc{\'\i}a-Sedano}, {Garofalo}, {Garralda}, {Gavel}, {Gavras}, {Gerssen},
  {Geyer}, {Giacobbe}, {Gilmore}, {Girona}, {Giuffrida}, {Glass}, {Gomes},
  {Granvik}, {Gueguen}, {Guerrier}, {Guiraud}, {Guti{\'e}rrez-S{\'a}nchez},
  {Haigron}, {Hatzidimitriou}, {Hauser}, {Haywood}, {Heiter}, {Helmi}, {Heu},
  {Hilger}, {Hobbs}, {Hofmann}, {Holland}, {Huckle}, {Hypki}, {Icardi},
  {Jan{\ss}en}, {Jevardat de Fombelle}, {Jonker}, {Juh{\'a}sz}, {Julbe},
  {Karampelas}, {Kewley}, {Klar}, {Kochoska}, {Kohley}, {Kolenberg},
  {Kontizas}, {Kontizas}, {Koposov}, {Kordopatis}, {Kostrzewa-Rutkowska},
  {Koubsky}, {Lambert}, {Lanza}, {Lasne}, {Lavigne}, {Le Fustec}, {Le
  Poncin-Lafitte}, {Lebreton}, {Leccia}, {Leclerc}, {Lecoeur-Taibi},
  {Lenhardt}, {Leroux}, {Liao}, {Licata}, {Lindstr{\o}m}, {Lister}, {Livanou},
  {Lobel}, {L{\'o}pez}, {Managau}, {Mann}, {Mantelet}, {Marchal}, {Marchant},
  {Marconi}, {Marinoni}, {Marschalk{\'o}}, {Marshall}, {Martino}, {Marton},
  {Mary}, {Massari}, {Matijevi{\v{c}}}, {Mazeh}, {McMillan}, {Messina},
  {Michalik}, {Millar}, {Molina}, {Molinaro}, {Moln{\'a}r}, {Montegriffo},
  {Mor}, {Morbidelli}, {Morel}, {Morris}, {Mulone}, {Muraveva}, {Musella},
  {Nelemans}, {Nicastro}, {Noval}, {O'Mullane}, {Ord{\'e}novic},
  {Ord{\'o}{\~n}ez-Blanco}, {Osborne}, {Pagani}, {Pagano}, {Pailler},
  {Palacin}, {Palaversa}, {Panahi}, {Pawlak}, {Piersimoni}, {Pineau}, {Plachy},
  {Plum}, {Poggio}, {Poujoulet}, {Pr{\v{s}}a}, {Pulone}, {Racero}, {Ragaini},
  {Rambaux}, {Ramos-Lerate}, {Regibo}, {Reyl{\'e}}, {Riclet}, {Ripepi}, {Riva},
  {Rivard}, {Rixon}, {Roegiers}, {Roelens}, {Romero-G{\'o}mez}, {Rowell},
  {Royer}, {Ruiz-Dern}, {Sadowski}, {Sagrist{\`a} Sell{\'e}s}, {Sahlmann},
  {Salgado}, {Salguero}, {Sanna}, {Santana-Ros}, {Sarasso}, {Savietto},
  {Schultheis}, {Sciacca}, {Segol}, {Segovia}, {S{\'e}gransan}, {Shih},
  {Siltala}, {Silva}, {Smart}, {Smith}, {Solano}, {Solitro}, {Sordo}, {Soria
  Nieto}, {Souchay}, {Spagna}, {Spoto}, {Stampa}, {Steele},
  {Steidelm{\"u}ller}, {Stephenson}, {Stoev}, {Suess}, {Surdej}, {Szabados},
  {Szegedi-Elek}, {Tapiador}, {Taris}, {Tauran}, {Taylor}, {Teixeira},
  {Terrett}, {Teyssand ier}, {Thuillot}, {Titarenko}, {Torra Clotet}, {Turon},
  {Ulla}, {Utrilla}, {Uzzi}, {Vaillant}, {Valentini}, {Valette}, {van Elteren},
  {Van Hemelryck}, {van Leeuwen}, {Vaschetto}, {Vecchiato}, {Veljanoski},
  {Viala}, {Vicente}, {Vogt}, {von Essen}, {Voss}, {Votruba}, {Voutsinas},
  {Walmsley}, {Weiler}, {Wertz}, {Wevers}, {Wyrzykowski}, {Yoldas},
  {{\v{Z}}erjal}, {Ziaeepour}, {Zorec}, {Zschocke}, {Zucker}, {Zurbach}, \&
  {Zwitter}}]{gaia2018}
{Gaia Collaboration}, {Brown}, A.~G.~A., {Vallenari}, A., {et~al.} 2018, \aap,
  616, A1, \dodoi{10.1051/0004-6361/201833051}

\bibitem[{{Gaia Collaboration} {et~al.}(2021){Gaia Collaboration}, {Brown},
  {Vallenari}, {Prusti}, {de Bruijne}, {Babusiaux}, {Biermann}, {Creevey},
  {Evans}, {Eyer}, {Hutton}, {Jansen}, {Jordi}, {Klioner}, {Lammers},
  {Lindegren}, {Luri}, {Mignard}, {Panem}, {Pourbaix}, {Randich}, {Sartoretti},
  {Soubiran}, {Walton}, {Arenou}, {Bailer-Jones}, {Bastian}, {Cropper},
  {Drimmel}, {Katz}, {Lattanzi}, {van Leeuwen}, {Bakker}, {Cacciari},
  {Casta{\~n}eda}, {De Angeli}, {Ducourant}, {Fabricius}, {Fouesneau},
  {Fr{\'e}mat}, {Guerra}, {Guerrier}, {Guiraud}, {Jean-Antoine Piccolo},
  {Masana}, {Messineo}, {Mowlavi}, {Nicolas}, {Nienartowicz}, {Pailler},
  {Panuzzo}, {Riclet}, {Roux}, {Seabroke}, {Sordo}, {Tanga}, {Th{\'e}venin},
  {Gracia-Abril}, {Portell}, {Teyssier}, {Altmann}, {Andrae}, {Bellas-Velidis},
  {Benson}, {Berthier}, {Blomme}, {Brugaletta}, {Burgess}, {Busso}, {Carry},
  {Cellino}, {Cheek}, {Clementini}, {Damerdji}, {Davidson}, {Delchambre},
  {Dell'Oro}, {Fern{\'a}ndez-Hern{\'a}ndez}, {Galluccio}, {Garc{\'\i}a-Lario},
  {Garcia-Reinaldos}, {Gonz{\'a}lez-N{\'u}{\~n}ez}, {Gosset}, {Haigron},
  {Halbwachs}, {Hambly}, {Harrison}, {Hatzidimitriou}, {Heiter},
  {Hern{\'a}ndez}, {Hestroffer}, {Hodgkin}, {Holl}, {Jan{\ss}en}, {Jevardat de
  Fombelle}, {Jordan}, {Krone-Martins}, {Lanzafame}, {L{\"o}ffler}, {Lorca},
  {Manteiga}, {Marchal}, {Marrese}, {Moitinho}, {Mora}, {Muinonen}, {Osborne},
  {Pancino}, {Pauwels}, {Petit}, {Recio-Blanco}, {Richards}, {Riello},
  {Rimoldini}, {Robin}, {Roegiers}, {Rybizki}, {Sarro}, {Siopis}, {Smith},
  {Sozzetti}, {Ulla}, {Utrilla}, {van Leeuwen}, {van Reeven}, {Abbas}, {Abreu
  Aramburu}, {Accart}, {Aerts}, {Aguado}, {Ajaj}, {Altavilla}, {{\'A}lvarez},
  {{\'A}lvarez Cid-Fuentes}, {Alves}, {Anderson}, {Anglada Varela}, {Antoja},
  {Audard}, {Baines}, {Baker}, {Balaguer-N{\'u}{\~n}ez}, {Balbinot}, {Balog},
  {Barache}, {Barbato}, {Barros}, {Barstow}, {Bartolom{\'e}}, {Bassilana},
  {Bauchet}, {Baudesson-Stella}, {Becciani}, {Bellazzini}, {Bernet}, {Bertone},
  {Bianchi}, {Blanco-Cuaresma}, {Boch}, {Bombrun}, {Bossini}, {Bouquillon},
  {Bragaglia}, {Bramante}, {Breedt}, {Bressan}, {Brouillet}, {Bucciarelli},
  {Burlacu}, {Busonero}, {Butkevich}, {Buzzi}, {Caffau}, {Cancelliere},
  {C{\'a}novas}, {Cantat-Gaudin}, {Carballo}, {Carlucci}, {Carnerero},
  {Carrasco}, {Casamiquela}, {Castellani}, {Castro-Ginard}, {Castro Sampol},
  {Chaoul}, {Charlot}, {Chemin}, {Chiavassa}, {Cioni}, {Comoretto}, {Cooper},
  {Cornez}, {Cowell}, {Crifo}, {Crosta}, {Crowley}, {Dafonte}, {Dapergolas},
  {David}, {David}, {de Laverny}, {De Luise}, {De March}, {De Ridder}, {de
  Souza}, {de Teodoro}, {de Torres}, {del Peloso}, {del Pozo}, {Delbo},
  {Delgado}, {Delgado}, {Delisle}, {Di Matteo}, {Diakite}, {Diener},
  {Distefano}, {Dolding}, {Eappachen}, {Edvardsson}, {Enke}, {Esquej}, {Fabre},
  {Fabrizio}, {Faigler}, {Fedorets}, {Fernique}, {Fienga}, {Figueras},
  {Fouron}, {Fragkoudi}, {Fraile}, {Franke}, {Gai}, {Garabato},
  {Garcia-Gutierrez}, {Garc{\'\i}a-Torres}, {Garofalo}, {Gavras}, {Gerlach},
  {Geyer}, {Giacobbe}, {Gilmore}, {Girona}, {Giuffrida}, {Gomel}, {Gomez},
  {Gonzalez-Santamaria}, {Gonz{\'a}lez-Vidal}, {Granvik},
  {Guti{\'e}rrez-S{\'a}nchez}, {Guy}, {Hauser}, {Haywood}, {Helmi}, {Hidalgo},
  {Hilger}, {H{\l}adczuk}, {Hobbs}, {Holland}, {Huckle}, {Jasniewicz},
  {Jonker}, {Juaristi Campillo}, {Julbe}, {Karbevska}, {Kervella}, {Khanna},
  {Kochoska}, {Kontizas}, {Kordopatis}, {Korn}, {Kostrzewa-Rutkowska},
  {Kruszy{\'n}ska}, {Lambert}, {Lanza}, {Lasne}, {Le Campion}, {Le Fustec},
  {Lebreton}, {Lebzelter}, {Leccia}, {Leclerc}, {Lecoeur-Taibi}, {Liao},
  {Licata}, {Lindstr{\o}m}, {Lister}, {Livanou}, {Lobel}, {Madrero Pardo},
  {Managau}, {Mann}, {Marchant}, {Marconi}, {Marcos Santos}, {Marinoni},
  {Marocco}, {Marshall}, {Martin Polo}, {Mart{\'\i}n-Fleitas}, {Masip},
  {Massari}, {Mastrobuono-Battisti}, {Mazeh}, {McMillan}, {Messina},
  {Michalik}, {Millar}, {Mints}, {Molina}, {Molinaro}, {Moln{\'a}r},
  {Montegriffo}, {Mor}, {Morbidelli}, {Morel}, {Morris}, {Mulone}, {Munoz},
  {Muraveva}, {Murphy}, {Musella}, {Noval}, {Ord{\'e}novic}, {Orr{\`u}},
  {Osinde}, {Pagani}, {Pagano}, {Palaversa}, {Palicio}, {Panahi}, {Pawlak},
  {Pe{\~n}alosa Esteller}, {Penttil{\"a}}, {Piersimoni}, {Pineau}, {Plachy},
  {Plum}, {Poggio}, {Poretti}, {Poujoulet}, {Pr{\v{s}}a}, {Pulone}, {Racero},
  {Ragaini}, {Rainer}, {Raiteri}, {Rambaux}, {Ramos}, {Ramos-Lerate}, {Re
  Fiorentin}, {Regibo}, {Reyl{\'e}}, {Ripepi}, {Riva}, {Rixon}, {Robichon},
  {Robin}, {Roelens}, {Rohrbasser}, {Romero-G{\'o}mez}, {Rowell}, {Royer},
  {Rybicki}, {Sadowski}, {Sagrist{\`a} Sell{\'e}s}, {Sahlmann}, {Salgado},
  {Salguero}, {Samaras}, {Sanchez Gimenez}, {Sanna}, {Santove{\~n}a},
  {Sarasso}, {Schultheis}, {Sciacca}, {Segol}, {Segovia}, {S{\'e}gransan},
  {Semeux}, {Shahaf}, {Siddiqui}, {Siebert}, {Siltala}, {Slezak}, {Smart},
  {Solano}, {Solitro}, {Souami}, {Souchay}, {Spagna}, {Spoto}, {Steele},
  {Steidelm{\"u}ller}, {Stephenson}, {S{\"u}veges}, {Szabados}, {Szegedi-Elek},
  {Taris}, {Tauran}, {Taylor}, {Teixeira}, {Thuillot}, {Tonello}, {Torra},
  {Torra}, {Turon}, {Unger}, {Vaillant}, {van Dillen}, {Vanel}, {Vecchiato},
  {Viala}, {Vicente}, {Voutsinas}, {Weiler}, {Wevers}, {Wyrzykowski}, {Yoldas},
  {Yvard}, {Zhao}, {Zorec}, {Zucker}, {Zurbach}, \& {Zwitter}}]{gaia2021}
---. 2021, \aap, 649, A1, \dodoi{10.1051/0004-6361/202039657}

\bibitem[{{Gaia Collaboration} {et~al.}(2023){Gaia Collaboration}, {Vallenari},
  {Brown}, {Prusti}, {de Bruijne}, {Arenou}, {Babusiaux}, {Biermann},
  {Creevey}, {Ducourant}, {Evans}, {Eyer}, {Guerra}, {Hutton}, {Jordi},
  {Klioner}, {Lammers}, {Lindegren}, {Luri}, {Mignard}, {Panem}, {Pourbaix},
  {Randich}, {Sartoretti}, {Soubiran}, {Tanga}, {Walton}, {Bailer-Jones},
  {Bastian}, {Drimmel}, {Jansen}, {Katz}, {Lattanzi}, {van Leeuwen}, {Bakker},
  {Cacciari}, {Casta{\~n}eda}, {De Angeli}, {Fabricius}, {Fouesneau},
  {Fr{\'e}mat}, {Galluccio}, {Guerrier}, {Heiter}, {Masana}, {Messineo},
  {Mowlavi}, {Nicolas}, {Nienartowicz}, {Pailler}, {Panuzzo}, {Riclet}, {Roux},
  {Seabroke}, {Sordo}, {Th{\'e}venin}, {Gracia-Abril}, {Portell}, {Teyssier},
  {Altmann}, {Andrae}, {Audard}, {Bellas-Velidis}, {Benson}, {Berthier},
  {Blomme}, {Burgess}, {Busonero}, {Busso}, {C{\'a}novas}, {Carry}, {Cellino},
  {Cheek}, {Clementini}, {Damerdji}, {Davidson}, {de Teodoro}, {Nu{\~n}ez
  Campos}, {Delchambre}, {Dell'Oro}, {Esquej}, {Fern{\'a}ndez-Hern{\'a}ndez},
  {Fraile}, {Garabato}, {Garc{\'\i}a-Lario}, {Gosset}, {Haigron}, {Halbwachs},
  {Hambly}, {Harrison}, {Hern{\'a}ndez}, {Hestroffer}, {Hodgkin}, {Holl},
  {Jan{\ss}en}, {Jevardat de Fombelle}, {Jordan}, {Krone-Martins}, {Lanzafame},
  {L{\"o}ffler}, {Marchal}, {Marrese}, {Moitinho}, {Muinonen}, {Osborne},
  {Pancino}, {Pauwels}, {Recio-Blanco}, {Reyl{\'e}}, {Riello}, {Rimoldini},
  {Roegiers}, {Rybizki}, {Sarro}, {Siopis}, {Smith}, {Sozzetti}, {Utrilla},
  {van Leeuwen}, {Abbas}, {{\'A}brah{\'a}m}, {Abreu Aramburu}, {Aerts},
  {Aguado}, {Ajaj}, {Aldea-Montero}, {Altavilla}, {{\'A}lvarez}, {Alves},
  {Anders}, {Anderson}, {Anglada Varela}, {Antoja}, {Baines}, {Baker},
  {Balaguer-N{\'u}{\~n}ez}, {Balbinot}, {Balog}, {Barache}, {Barbato},
  {Barros}, {Barstow}, {Bartolom{\'e}}, {Bassilana}, {Bauchet}, {Becciani},
  {Bellazzini}, {Berihuete}, {Bernet}, {Bertone}, {Bianchi}, {Binnenfeld},
  {Blanco-Cuaresma}, {Blazere}, {Boch}, {Bombrun}, {Bossini}, {Bouquillon},
  {Bragaglia}, {Bramante}, {Breedt}, {Bressan}, {Brouillet}, {Brugaletta},
  {Bucciarelli}, {Burlacu}, {Butkevich}, {Buzzi}, {Caffau}, {Cancelliere},
  {Cantat-Gaudin}, {Carballo}, {Carlucci}, {Carnerero}, {Carrasco},
  {Casamiquela}, {Castellani}, {Castro-Ginard}, {Chaoul}, {Charlot}, {Chemin},
  {Chiaramida}, {Chiavassa}, {Chornay}, {Comoretto}, {Contursi}, {Cooper},
  {Cornez}, {Cowell}, {Crifo}, {Cropper}, {Crosta}, {Crowley}, {Dafonte},
  {Dapergolas}, {David}, {David}, {de Laverny}, {De Luise}, {De March}, {De
  Ridder}, {de Souza}, {de Torres}, {del Peloso}, {del Pozo}, {Delbo},
  {Delgado}, {Delisle}, {Demouchy}, {Dharmawardena}, {Di Matteo}, {Diakite},
  {Diener}, {Distefano}, {Dolding}, {Edvardsson}, {Enke}, {Fabre}, {Fabrizio},
  {Faigler}, {Fedorets}, {Fernique}, {Fienga}, {Figueras}, {Fournier},
  {Fouron}, {Fragkoudi}, {Gai}, {Garcia-Gutierrez}, {Garcia-Reinaldos},
  {Garc{\'\i}a-Torres}, {Garofalo}, {Gavel}, {Gavras}, {Gerlach}, {Geyer},
  {Giacobbe}, {Gilmore}, {Girona}, {Giuffrida}, {Gomel}, {Gomez},
  {Gonz{\'a}lez-N{\'u}{\~n}ez}, {Gonz{\'a}lez-Santamar{\'\i}a},
  {Gonz{\'a}lez-Vidal}, {Granvik}, {Guillout}, {Guiraud},
  {Guti{\'e}rrez-S{\'a}nchez}, {Guy}, {Hatzidimitriou}, {Hauser}, {Haywood},
  {Helmer}, {Helmi}, {Sarmiento}, {Hidalgo}, {Hilger}, {H{\l}adczuk}, {Hobbs},
  {Holland}, {Huckle}, {Jardine}, {Jasniewicz}, {Jean-Antoine Piccolo},
  {Jim{\'e}nez-Arranz}, {Jorissen}, {Juaristi Campillo}, {Julbe}, {Karbevska},
  {Kervella}, {Khanna}, {Kontizas}, {Kordopatis}, {Korn}, {K{\'o}sp{\'a}l},
  {Kostrzewa-Rutkowska}, {Kruszy{\'n}ska}, {Kun}, {Laizeau}, {Lambert},
  {Lanza}, {Lasne}, {Le Campion}, {Lebreton}, {Lebzelter}, {Leccia}, {Leclerc},
  {Lecoeur-Taibi}, {Liao}, {Licata}, {Lindstr{\o}m}, {Lister}, {Livanou},
  {Lobel}, {Lorca}, {Loup}, {Madrero Pardo}, {Magdaleno Romeo}, {Managau},
  {Mann}, {Manteiga}, {Marchant}, {Marconi}, {Marcos}, {Marcos Santos},
  {Mar{\'\i}n Pina}, {Marinoni}, {Marocco}, {Marshall}, {Martin Polo},
  {Mart{\'\i}n-Fleitas}, {Marton}, {Mary}, {Masip}, {Massari},
  {Mastrobuono-Battisti}, {Mazeh}, {McMillan}, {Messina}, {Michalik}, {Millar},
  {Mints}, {Molina}, {Molinaro}, {Moln{\'a}r}, {Monari}, {Mongui{\'o}},
  {Montegriffo}, {Montero}, {Mor}, {Mora}, {Morbidelli}, {Morel}, {Morris},
  {Muraveva}, {Murphy}, {Musella}, {Nagy}, {Noval}, {Oca{\~n}a}, {Ogden},
  {Ordenovic}, {Osinde}, {Pagani}, {Pagano}, {Palaversa}, {Palicio},
  {Pallas-Quintela}, {Panahi}, {Payne-Wardenaar}, {Pe{\~n}alosa Esteller},
  {Penttil{\"a}}, {Pichon}, {Piersimoni}, {Pineau}, {Plachy}, {Plum}, {Poggio},
  {Pr{\v{s}}a}, {Pulone}, {Racero}, {Ragaini}, {Rainer}, {Raiteri}, {Rambaux},
  {Ramos}, {Ramos-Lerate}, {Re Fiorentin}, {Regibo}, {Richards}, {Rios Diaz},
  {Ripepi}, {Riva}, {Rix}, {Rixon}, {Robichon}, {Robin}, {Robin}, {Roelens},
  {Rogues}, {Rohrbasser}, {Romero-G{\'o}mez}, {Rowell}, {Royer}, {Ruz Mieres},
  {Rybicki}, {Sadowski}, {S{\'a}ez N{\'u}{\~n}ez}, {Sagrist{\`a} Sell{\'e}s},
  {Sahlmann}, {Salguero}, {Samaras}, {Sanchez Gimenez}, {Sanna},
  {Santove{\~n}a}, {Sarasso}, {Schultheis}, {Sciacca}, {Segol}, {Segovia},
  {S{\'e}gransan}, {Semeux}, {Shahaf}, {Siddiqui}, {Siebert}, {Siltala},
  {Silvelo}, {Slezak}, {Slezak}, {Smart}, {Snaith}, {Solano}, {Solitro},
  {Souami}, {Souchay}, {Spagna}, {Spina}, {Spoto}, {Steele},
  {Steidelm{\"u}ller}, {Stephenson}, {S{\"u}veges}, {Surdej}, {Szabados},
  {Szegedi-Elek}, {Taris}, {Taylor}, {Teixeira}, {Tolomei}, {Tonello}, {Torra},
  {Torra}, {Torralba Elipe}, {Trabucchi}, {Tsounis}, {Turon}, {Ulla}, {Unger},
  {Vaillant}, {van Dillen}, {van Reeven}, {Vanel}, {Vecchiato}, {Viala},
  {Vicente}, {Voutsinas}, {Weiler}, {Wevers}, {Wyrzykowski}, {Yoldas}, {Yvard},
  {Zhao}, {Zorec}, {Zucker}, \& {Zwitter}}]{gaia2023}
{Gaia Collaboration}, {Vallenari}, A., {Brown}, A.~G.~A., {et~al.} 2023, \aap,
  674, A1, \dodoi{10.1051/0004-6361/202243940}

\bibitem[{{Gaudi} {et~al.}(2020){Gaudi}, {Seager}, {Mennesson}, {Kiessling},
  {Warfield}, {Cahoy}, {Clarke}, {Domagal-Goldman}, {Feinberg}, {Guyon},
  {Kasdin}, {Mawet}, {Plavchan}, {Robinson}, {Rogers}, {Scowen}, {Somerville},
  {Stapelfeldt}, {Stark}, {Stern}, {Turnbull}, {Amini}, {Kuan}, {Martin},
  {Morgan}, {Redding}, {Stahl}, {Webb}, {Alvarez-Salazar}, {Arnold}, {Arya},
  {Balasubramanian}, {Baysinger}, {Bell}, {Below}, {Benson}, {Blais}, {Booth},
  {Bourgeois}, {Bradford}, {Brewer}, {Brooks}, {Cady}, {Caldwell}, {Calvet},
  {Carr}, {Chan}, {Cormarkovic}, {Coste}, {Cox}, {Danner}, {Davis}, {Dewell},
  {Dorsett}, {Dunn}, {East}, {Effinger}, {Eng}, {Freebury}, {Garcia}, {Gaskin},
  {Greene}, {Hennessy}, {Hilgemann}, {Hood}, {Holota}, {Howe}, {Huang}, {Hull},
  {Hunt}, {Hurd}, {Johnson}, {Kissil}, {Knight}, {Kolenz}, {Kraus}, {Krist},
  {Li}, {Lisman}, {Mandic}, {Mann}, {Marchen}, {Marrese-Reading}, {McCready},
  {McGown}, {Missun}, {Miyaguchi}, {Moore}, {Nemati}, {Nikzad}, {Nissen},
  {Novicki}, {Perrine}, {Pineda}, {Polanco}, {Putnam}, {Qureshi}, {Richards},
  {Eldorado Riggs}, {Rodgers}, {Rud}, {Saini}, {Scalisi}, {Scharf}, {Schulz},
  {Serabyn}, {Sigrist}, {Sikkia}, {Singleton}, {Shaklan}, {Smith}, {Southerd},
  {Stahl}, {Steeves}, {Sturges}, {Sullivan}, {Tang}, {Taras}, {Tesch},
  {Therrell}, {Tseng}, {Valente}, {Van Buren}, {Villalvazo}, {Warwick}, {Webb},
  {Westerhoff}, {Wofford}, {Wu}, {Woo}, {Wood}, {Ziemer}, {Arney}, {Anderson},
  {Ma{\'\i}z-Apell{\'a}niz}, {Bartlett}, {Belikov}, {Bendek}, {Cenko},
  {Douglas}, {Dulz}, {Evans}, {Faramaz}, {Feng}, {Ferguson}, {Follette},
  {Ford}, {Garc{\'\i}a}, {Geha}, {Gelino}, {G{\"o}tberg}, {Hildebrand t}, {Hu},
  {Jahnke}, {Kennedy}, {Kreidberg}, {Isella}, {Lopez}, {Marchis}, {Macri},
  {Marley}, {Matzko}, {Mazoyer}, {McCandliss}, {Meshkat}, {Mordasini},
  {Morris}, {Nielsen}, {Newman}, {Petigura}, {Postman}, {Reines}, {Roberge},
  {Roederer}, {Ruane}, {Schwieterman}, {Sirbu}, {Spalding}, {Teplitz},
  {Tumlinson}, {Turner}, {Werk}, {Wofford}, {Wyatt}, {Young}, \&
  {Zellem}}]{gaudi2020}
{Gaudi}, B.~S., {Seager}, S., {Mennesson}, B., {et~al.} 2020, arXiv e-prints,
  arXiv:2001.06683.
\newblock \doarXiv{2001.06683}

\bibitem[{{Giles} {et~al.}(2017){Giles}, {Collier Cameron}, \&
  {Haywood}}]{giles2017}
{Giles}, H. A.~C., {Collier Cameron}, A., \& {Haywood}, R.~D. 2017, \mnras,
  472, 1618, \dodoi{10.1093/mnras/stx1931}

\bibitem[{{Gomes da Silva} {et~al.}(2022){Gomes da Silva}, {Bensabat},
  {Monteiro}, \& {Santos}}]{silva2022}
{Gomes da Silva}, J., {Bensabat}, A., {Monteiro}, T., \& {Santos}, N.~C. 2022,
  \aap, 668, A174, \dodoi{10.1051/0004-6361/202244595}

\bibitem[{{Gomes da Silva} {et~al.}(2018){Gomes da Silva}, {Figueira},
  {Santos}, \& {Faria}}]{gsilva2018}
{Gomes da Silva}, J., {Figueira}, P., {Santos}, N., \& {Faria}, J. 2018, The
  Journal of Open Source Software, 3, 667, \dodoi{10.21105/joss.00667}

\bibitem[{{Gomes da Silva} {et~al.}(2021){Gomes da Silva}, {Santos},
  {Adibekyan}, {Sousa}, {Campante}, {Figueira}, {Bossini}, {Delgado-Mena},
  {Monteiro}, {de Laverny}, {Recio-Blanco}, \& {Lovis}}]{gsilva2021}
{Gomes da Silva}, J., {Santos}, N.~C., {Adibekyan}, V., {et~al.} 2021, \aap,
  646, A77, \dodoi{10.1051/0004-6361/202039765}

\bibitem[{{Henry} \& {McCarthy}(1993)}]{henry1993}
{Henry}, T.~J., \& {McCarthy}, Jr., D.~W. 1993, \aj, 106, 773,
  \dodoi{10.1086/116685}

\bibitem[{{Hill} {et~al.}(2023){Hill}, {Bott}, {Dalba}, {Fetherolf}, {Kane},
  {Kopparapu}, {Li}, \& {Ostberg}}]{hill2023}
{Hill}, M.~L., {Bott}, K., {Dalba}, P.~A., {et~al.} 2023, \aj, 165, 34,
  \dodoi{10.3847/1538-3881/aca1c0}

\bibitem[{{Hill} {et~al.}(2018){Hill}, {Kane}, {Seperuelo Duarte}, {Kopparapu},
  {Gelino}, \& {Wittenmyer}}]{hill2018}
{Hill}, M.~L., {Kane}, S.~R., {Seperuelo Duarte}, E., {et~al.} 2018, \apj, 860,
  67, \dodoi{10.3847/1538-4357/aac384}

\bibitem[{{H{\o}g} {et~al.}(2000){H{\o}g}, {Fabricius}, {Makarov}, {Urban},
  {Corbin}, {Wycoff}, {Bastian}, {Schwekendiek}, \& {Wicenec}}]{hog2000}
{H{\o}g}, E., {Fabricius}, C., {Makarov}, V.~V., {et~al.} 2000, \aap, 355, L27

\bibitem[{{Horch} {et~al.}(2011){Horch}, {Gomez}, {Sherry}, {Howell}, {Ciardi},
  {Anderson}, \& {van Altena}}]{horch2011}
{Horch}, E.~P., {Gomez}, S.~C., {Sherry}, W.~H., {et~al.} 2011, \aj, 141, 45,
  \dodoi{10.1088/0004-6256/141/2/45}

\bibitem[{{Horner} {et~al.}(2020){Horner}, {Kane}, {Marshall}, {Dalba}, {Holt},
  {Wood}, {Maynard-Casely}, {Wittenmyer}, {Lykawka}, {Hill}, {Salmeron},
  {Bailey}, {L{\"o}hne}, {Agnew}, {Carter}, \& {Tylor}}]{horner2020b}
{Horner}, J., {Kane}, S.~R., {Marshall}, J.~P., {et~al.} 2020, \pasp, 132,
  102001, \dodoi{10.1088/1538-3873/ab8eb9}

\bibitem[{{Howard} \& {Fulton}(2016)}]{howard2016}
{Howard}, A.~W., \& {Fulton}, B.~J. 2016, \pasp, 128, 114401,
  \dodoi{10.1088/1538-3873/128/969/114401}

\bibitem[{{Howell} {et~al.}(2011){Howell}, {Everett}, {Sherry}, {Horch}, \&
  {Ciardi}}]{howell2011}
{Howell}, S.~B., {Everett}, M.~E., {Sherry}, W., {Horch}, E., \& {Ciardi},
  D.~R. 2011, \aj, 142, 19, \dodoi{10.1088/0004-6256/142/1/19}

\bibitem[{{Jenkins} {et~al.}(2016){Jenkins}, {Twicken}, {McCauliff},
  {Campbell}, {Sanderfer}, {Lung}, {Mansouri-Samani}, {Girouard}, {Tenenbaum},
  {Klaus}, {Smith}, {Caldwell}, {Chacon}, {Henze}, {Heiges}, {Latham},
  {Morgan}, {Swade}, {Rinehart}, \& {Vanderspek}}]{jenkins2016}
{Jenkins}, J.~M., {Twicken}, J.~D., {McCauliff}, S., {et~al.} 2016, in
  \procspie, Vol. 9913, Software and Cyberinfrastructure for Astronomy IV,
  99133E, \dodoi{10.1117/12.2233418}

\bibitem[{{Kane}(2013)}]{kane2013c}
{Kane}, S.~R. 2013, \apj, 766, 10, \dodoi{10.1088/0004-637X/766/1/10}

\bibitem[{{Kane}(2022)}]{kane2022b}
---. 2022, Nature Astronomy, 6, 420, \dodoi{10.1038/s41550-022-01626-x}

\bibitem[{{Kane} \& {Gelino}(2012)}]{kane2012a}
{Kane}, S.~R., \& {Gelino}, D.~M. 2012, \pasp, 124, 323, \dodoi{10.1086/665271}

\bibitem[{{Kane} {et~al.}(2007){Kane}, {Schneider}, \& {Ge}}]{kane2007a}
{Kane}, S.~R., {Schneider}, D.~P., \& {Ge}, J. 2007, \mnras, 377, 1610,
  \dodoi{10.1111/j.1365-2966.2007.11722.x}

\bibitem[{{Kane} {et~al.}(2014){Kane}, {Howell}, {Horch}, {Feng}, {Hinkel},
  {Ciardi}, {Everett}, {Howard}, \& {Wright}}]{kane2014c}
{Kane}, S.~R., {Howell}, S.~B., {Horch}, E.~P., {et~al.} 2014, \apj, 785, 93,
  \dodoi{10.1088/0004-637X/785/2/93}

\bibitem[{{Kane} {et~al.}(2016{\natexlab{a}}){Kane}, {Thirumalachari}, {Henry},
  {Hinkel}, {Jensen}, {Boyajian}, {Fischer}, {Howard}, {Isaacson}, \&
  {Wright}}]{kane2016a}
{Kane}, S.~R., {Thirumalachari}, B., {Henry}, G.~W., {et~al.}
  2016{\natexlab{a}}, \apjl, 820, L5, \dodoi{10.3847/2041-8205/820/1/L5}

\bibitem[{{Kane} {et~al.}(2016{\natexlab{b}}){Kane}, {Hill}, {Kasting},
  {Kopparapu}, {Quintana}, {Barclay}, {Batalha}, {Borucki}, {Ciardi},
  {Haghighipour}, {Hinkel}, {Kaltenegger}, {Selsis}, \& {Torres}}]{kane2016c}
{Kane}, S.~R., {Hill}, M.~L., {Kasting}, J.~F., {et~al.} 2016{\natexlab{b}},
  \apj, 830, 1, \dodoi{10.3847/0004-637X/830/1/1}

\bibitem[{{Kane} {et~al.}(2019){Kane}, {Dalba}, {Li}, {Horch}, {Hirsch},
  {Horner}, {Wittenmyer}, {Howell}, {Everett}, {Butler}, {Tinney}, {Carter},
  {Wright}, {Jones}, {Bailey}, \& {O'Toole}}]{kane2019b}
{Kane}, S.~R., {Dalba}, P.~A., {Li}, Z., {et~al.} 2019, \aj, 157, 252,
  \dodoi{10.3847/1538-3881/ab1ddf}

\bibitem[{{Kane} {et~al.}(2021){Kane}, {Arney}, {Byrne}, {Dalba}, {Desch},
  {Horner}, {Izenberg}, {Mandt}, {Meadows}, \& {Quick}}]{kane2021d}
{Kane}, S.~R., {Arney}, G.~N., {Byrne}, P.~K., {et~al.} 2021, Journal of
  Geophysical Research (Planets), 126, e06643, \dodoi{10.1002/jgre.v126.2}

\bibitem[{{Kasting} {et~al.}(1993){Kasting}, {Whitmire}, \&
  {Reynolds}}]{kasting1993a}
{Kasting}, J.~F., {Whitmire}, D.~P., \& {Reynolds}, R.~T. 1993, \icarus, 101,
  108, \dodoi{10.1006/icar.1993.1010}

\bibitem[{{Kipping}(2013)}]{kipping2013b}
{Kipping}, D.~M. 2013, \mnras, 434, L51, \dodoi{10.1093/mnrasl/slt075}

\bibitem[{{Kopparapu} {et~al.}(2014){Kopparapu}, {Ramirez}, {SchottelKotte},
  {Kasting}, {Domagal-Goldman}, \& {Eymet}}]{kopparapu2014}
{Kopparapu}, R.~K., {Ramirez}, R.~M., {SchottelKotte}, J., {et~al.} 2014, \apj,
  787, L29, \dodoi{10.1088/2041-8205/787/2/L29}

\bibitem[{{Kopparapu} {et~al.}(2013){Kopparapu}, {Ramirez}, {Kasting}, {Eymet},
  {Robinson}, {Mahadevan}, {Terrien}, {Domagal-Goldman}, {Meadows}, \&
  {Deshpande}}]{kopparapu2013a}
{Kopparapu}, R.~K., {Ramirez}, R., {Kasting}, J.~F., {et~al.} 2013, \apj, 765,
  131, \dodoi{10.1088/0004-637X/765/2/131}

\bibitem[{{Kopparapu} {et~al.}(2018){Kopparapu}, {H{\'e}brard}, {Belikov},
  {Batalha}, {Mulders}, {Stark}, {Teal}, {Domagal-Goldman}, \&
  {Mandell}}]{kopparapu2018}
{Kopparapu}, R.~K., {H{\'e}brard}, E., {Belikov}, R., {et~al.} 2018, \apj, 856,
  122, \dodoi{10.3847/1538-4357/aab205}

\bibitem[{{Laliotis} {et~al.}(2023){Laliotis}, {Burt}, {Mamajek}, {Li},
  {Perdelwitz}, {Zhao}, {Butler}, {Holden}, {Rosenthal}, {Fulton}, {Feng},
  {Kane}, {Bailey}, {Carter}, {Crane}, {Furlan}, {Gnilka}, {Howell},
  {Laughlin}, {Shectman}, {Teske}, {Tinney}, {Vogt}, {Wang}, \&
  {Wittenmyer}}]{laliotis2023}
{Laliotis}, K., {Burt}, J.~A., {Mamajek}, E.~E., {et~al.} 2023, \aj, 165, 176,
  \dodoi{10.3847/1538-3881/acc067}

\bibitem[{{Li} {et~al.}(2021){Li}, {Hildebrandt}, {Kane}, {Zimmerman},
  {Girard}, {Gonzalez-Quiles}, \& {Turnbull}}]{li2021}
{Li}, Z., {Hildebrandt}, S.~R., {Kane}, S.~R., {et~al.} 2021, \aj, 162, 9,
  \dodoi{10.3847/1538-3881/abf831}

\bibitem[{{Li} {et~al.}(2022){Li}, {Kane}, {Dalba}, {Howard}, \&
  {Isaacson}}]{li2022}
{Li}, Z., {Kane}, S.~R., {Dalba}, P.~A., {Howard}, A.~W., \& {Isaacson}, H.~T.
  2022, \aj, 164, 163, \dodoi{10.3847/1538-3881/ac8d63}

\bibitem[{{Lindegren} {et~al.}(2021{\natexlab{a}}){Lindegren}, {Klioner},
  {Hern{\'a}ndez}, {Bombrun}, {Ramos-Lerate}, {Steidelm{\"u}ller}, {Bastian},
  {Biermann}, {de Torres}, {Gerlach}, {Geyer}, {Hilger}, {Hobbs}, {Lammers},
  {McMillan}, {Stephenson}, {Casta{\~n}eda}, {Davidson}, {Fabricius},
  {Gracia-Abril}, {Portell}, {Rowell}, {Teyssier}, {Torra}, {Bartolom{\'e}},
  {Clotet}, {Garralda}, {Gonz{\'a}lez-Vidal}, {Torra}, {Abbas}, {Altmann},
  {Anglada Varela}, {Balaguer-N{\'u}{\~n}ez}, {Balog}, {Barache}, {Becciani},
  {Bernet}, {Bertone}, {Bianchi}, {Bouquillon}, {Brown}, {Bucciarelli},
  {Busonero}, {Butkevich}, {Buzzi}, {Cancelliere}, {Carlucci}, {Charlot},
  {Cioni}, {Crosta}, {Crowley}, {del Peloso}, {del Pozo}, {Drimmel}, {Esquej},
  {Fienga}, {Fraile}, {Gai}, {Garcia-Reinaldos}, {Guerra}, {Hambly}, {Hauser},
  {Jan{\ss}en}, {Jordan}, {Kostrzewa-Rutkowska}, {Lattanzi}, {Liao}, {Licata},
  {Lister}, {L{\"o}ffler}, {Marchant}, {Masip}, {Mignard}, {Mints}, {Molina},
  {Mora}, {Morbidelli}, {Murphy}, {Pagani}, {Panuzzo}, {Pe{\~n}alosa Esteller},
  {Poggio}, {Re Fiorentin}, {Riva}, {Sagrist{\`a} Sell{\'e}s}, {Sanchez
  Gimenez}, {Sarasso}, {Sciacca}, {Siddiqui}, {Smart}, {Souami}, {Spagna},
  {Steele}, {Taris}, {Utrilla}, {van Reeven}, \& {Vecchiato}}]{lindegren2021a}
{Lindegren}, L., {Klioner}, S.~A., {Hern{\'a}ndez}, J., {et~al.}
  2021{\natexlab{a}}, \aap, 649, A2, \dodoi{10.1051/0004-6361/202039709}

\bibitem[{{Lindegren} {et~al.}(2021{\natexlab{b}}){Lindegren}, {Bastian},
  {Biermann}, {Bombrun}, {de Torres}, {Gerlach}, {Geyer}, {Hern{\'a}ndez},
  {Hilger}, {Hobbs}, {Klioner}, {Lammers}, {McMillan}, {Ramos-Lerate},
  {Steidelm{\"u}ller}, {Stephenson}, \& {van Leeuwen}}]{lindegren2021b}
{Lindegren}, L., {Bastian}, U., {Biermann}, M., {et~al.} 2021{\natexlab{b}},
  \aap, 649, A4, \dodoi{10.1051/0004-6361/202039653}

\bibitem[{{Lissauer} {et~al.}(2011){Lissauer}, {Ragozzine}, {Fabrycky},
  {Steffen}, {Ford}, {Jenkins}, {Shporer}, {Holman}, {Rowe}, {Quintana},
  {Batalha}, {Borucki}, {Bryson}, {Caldwell}, {Carter}, {Ciardi}, {Dunham},
  {Fortney}, {Gautier}, {Howell}, {Koch}, {Latham}, {Marcy}, {Morehead}, \&
  {Sasselov}}]{lissauer2011b}
{Lissauer}, J.~J., {Ragozzine}, D., {Fabrycky}, D.~C., {et~al.} 2011, \apjs,
  197, 8, \dodoi{10.1088/0067-0049/197/1/8}

\bibitem[{{Lo Curto} {et~al.}(2015){Lo Curto}, {Pepe}, {Avila}, {Boffin},
  {Bovay}, {Chazelas}, {Coffinet}, {Fleury}, {Hughes}, {Lovis}, {Maire},
  {Manescau}, {Pasquini}, {Rihs}, {Sinclaire}, \& {Udry}}]{locurto2015}
{Lo Curto}, G., {Pepe}, F., {Avila}, G., {et~al.} 2015, The Messenger, 162, 9

\bibitem[{{Lomb}(1976)}]{lomb1976}
{Lomb}, N.~R. 1976, \apss, 39, 447, \dodoi{10.1007/BF00648343}

\bibitem[{{Lubin} {et~al.}(2021){Lubin}, {Robertson}, {Stefansson}, {Ninan},
  {Mahadevan}, {Endl}, {Ford}, {Wright}, {Beard}, {Bender}, {Cochran},
  {Diddams}, {Fredrick}, {Halverson}, {Kanodia}, {Metcalf}, {Ramsey}, {Roy},
  {Schwab}, \& {Terrien}}]{lubin2021}
{Lubin}, J., {Robertson}, P., {Stefansson}, G., {et~al.} 2021, \aj, 162, 61,
  \dodoi{10.3847/1538-3881/ac0057}

\bibitem[{{Mamajek} \& {Hillenbrand}(2008)}]{mamajek2008}
{Mamajek}, E.~E., \& {Hillenbrand}, L.~A. 2008, \apj, 687, 1264,
  \dodoi{10.1086/591785}

\bibitem[{{Marley} {et~al.}(1999){Marley}, {Gelino}, {Stephens}, {Lunine}, \&
  {Freedman}}]{marley1999}
{Marley}, M.~S., {Gelino}, C., {Stephens}, D., {Lunine}, J.~I., \& {Freedman},
  R. 1999, \apj, 513, 879, \dodoi{10.1086/306881}

\bibitem[{{Mayor} {et~al.}(2011){Mayor}, {Marmier}, {Lovis}, {Udry},
  {S{\'e}gransan}, {Pepe}, {Benz}, {Bertaux}, {Bouchy}, {Dumusque}, {Lo Curto},
  {Mordasini}, {Queloz}, \& {Santos}}]{mayor2011}
{Mayor}, M., {Marmier}, M., {Lovis}, C., {et~al.} 2011, arXiv e-prints,
  arXiv:1109.2497.
\newblock \doarXiv{1109.2497}

\bibitem[{{Meunier} {et~al.}(2010){Meunier}, {Desort}, \&
  {Lagrange}}]{meunier2010}
{Meunier}, N., {Desort}, M., \& {Lagrange}, A.~M. 2010, \aap, 512, A39,
  \dodoi{10.1051/0004-6361/200913551}

\bibitem[{{Meunier} \& {Lagrange}(2019)}]{meunier2019b}
{Meunier}, N., \& {Lagrange}, A.~M. 2019, \aap, 629, A42,
  \dodoi{10.1051/0004-6361/201935651}

\bibitem[{{Mishra} {et~al.}(2023{\natexlab{a}}){Mishra}, {Alibert}, {Udry}, \&
  {Mordasini}}]{mishra2023a}
{Mishra}, L., {Alibert}, Y., {Udry}, S., \& {Mordasini}, C. 2023{\natexlab{a}},
  \aap, 670, A68, \dodoi{10.1051/0004-6361/202243751}

\bibitem[{{Mishra} {et~al.}(2023{\natexlab{b}}){Mishra}, {Alibert}, {Udry}, \&
  {Mordasini}}]{mishra2023b}
---. 2023{\natexlab{b}}, \aap, 670, A69, \dodoi{10.1051/0004-6361/202244705}

\bibitem[{{National Academies of Sciences, Engineering, and
  Medicine}(2021)}]{nas2021}
{National Academies of Sciences, Engineering, and Medicine}. 2021, {Pathways to
  Discovery in Astronomy and Astrophysics for the 2020s},
  \dodoi{10.17226/26141}

\bibitem[{{Newman} {et~al.}(2023){Newman}, {Plavchan}, {Burt}, {Teske},
  {Mamajek}, {Leifer}, {Gaudi}, {Blackwood}, \& {Morgan}}]{newman2023}
{Newman}, P.~D., {Plavchan}, P., {Burt}, J.~A., {et~al.} 2023, \aj, 165, 151,
  \dodoi{10.3847/1538-3881/acad07}

\bibitem[{{Nielsen} {et~al.}(2019){Nielsen}, {De Rosa}, {Macintosh}, {Wang},
  {Ruffio}, {Chiang}, {Marley}, {Saumon}, {Savransky}, {Ammons}, {Bailey},
  {Barman}, {Blain}, {Bulger}, {Burrows}, {Chilcote}, {Cotten}, {Czekala},
  {Doyon}, {Duch{\^e}ne}, {Esposito}, {Fabrycky}, {Fitzgerald}, {Follette},
  {Fortney}, {Gerard}, {Goodsell}, {Graham}, {Greenbaum}, {Hibon}, {Hinkley},
  {Hirsch}, {Hom}, {Hung}, {Dawson}, {Ingraham}, {Kalas}, {Konopacky},
  {Larkin}, {Lee}, {Lin}, {Maire}, {Marchis}, {Marois}, {Metchev},
  {Millar-Blanchaer}, {Morzinski}, {Oppenheimer}, {Palmer}, {Patience},
  {Perrin}, {Poyneer}, {Pueyo}, {Rafikov}, {Rajan}, {Rameau}, {Rantakyr{\"o}},
  {Ren}, {Schneider}, {Sivaramakrishnan}, {Song}, {Soummer}, {Tallis},
  {Thomas}, {Ward-Duong}, \& {Wolff}}]{nielsen2019c}
{Nielsen}, E.~L., {De Rosa}, R.~J., {Macintosh}, B., {et~al.} 2019, \aj, 158,
  13, \dodoi{10.3847/1538-3881/ab16e9}

\bibitem[{{Noyes} {et~al.}(1984){Noyes}, {Hartmann}, {Baliunas}, {Duncan}, \&
  {Vaughan}}]{noyes1984b}
{Noyes}, R.~W., {Hartmann}, L.~W., {Baliunas}, S.~L., {Duncan}, D.~K., \&
  {Vaughan}, A.~H. 1984, \apj, 279, 763, \dodoi{10.1086/161945}

\bibitem[{{Paxton} {et~al.}(2011){Paxton}, {Bildsten}, {Dotter}, {Herwig},
  {Lesaffre}, \& {Timmes}}]{paxton2011}
{Paxton}, B., {Bildsten}, L., {Dotter}, A., {et~al.} 2011, \apjs, 192, 3,
  \dodoi{10.1088/0067-0049/192/1/3}

\bibitem[{{Paxton} {et~al.}(2013){Paxton}, {Cantiello}, {Arras}, {Bildsten},
  {Brown}, {Dotter}, {Mankovich}, {Montgomery}, {Stello}, {Timmes}, \&
  {Townsend}}]{paxton2013}
{Paxton}, B., {Cantiello}, M., {Arras}, P., {et~al.} 2013, \apjs, 208, 4,
  \dodoi{10.1088/0067-0049/208/1/4}

\bibitem[{{Paxton} {et~al.}(2015){Paxton}, {Marchant}, {Schwab}, {Bauer},
  {Bildsten}, {Cantiello}, {Dessart}, {Farmer}, {Hu}, {Langer}, {Townsend},
  {Townsley}, \& {Timmes}}]{paxton2015}
{Paxton}, B., {Marchant}, P., {Schwab}, J., {et~al.} 2015, \apjs, 220, 15,
  \dodoi{10.1088/0067-0049/220/1/15}

\bibitem[{{Pecaut} \& {Mamajek}(2013)}]{pecaut2013}
{Pecaut}, M.~J., \& {Mamajek}, E.~E. 2013, \apjs, 208, 9,
  \dodoi{10.1088/0067-0049/208/1/9}

\bibitem[{{Pepe} {et~al.}(2000){Pepe}, {Mayor}, {Delabre}, {Kohler}, {Lacroix},
  {Queloz}, {Udry}, {Benz}, {Bertaux}, \& {Sivan}}]{pepe2000}
{Pepe}, F., {Mayor}, M., {Delabre}, B., {et~al.} 2000, in Society of
  Photo-Optical Instrumentation Engineers (SPIE) Conference Series, Vol. 4008,
  \procspie, ed. M.~{Iye} \& A.~F. {Moorwood}, 582--592,
  \dodoi{10.1117/12.395516}

\bibitem[{{Perdelwitz} {et~al.}(2023){Perdelwitz}, {Trifonov}, {Teklu},
  {Sreenivas}, \& {Tal-Or}}]{perdelwitz2023}
{Perdelwitz}, V., {Trifonov}, T., {Teklu}, J.~T., {Sreenivas}, K.~R., \&
  {Tal-Or}, L. 2023, arXiv e-prints, arXiv:2311.12438,
  \dodoi{10.48550/arXiv.2311.12438}

\bibitem[{{Pickles}(1998)}]{pickles1998}
{Pickles}, A.~J. 1998, \pasp, 110, 863, \dodoi{10.1086/316197}

\bibitem[{{Quanz} {et~al.}(2022){Quanz}, {Absil}, {Benz}, {Bonfils}, {Berger},
  {Defr{\`e}re}, {van Dishoeck}, {Ehrenreich}, {Fortney}, {Glauser},
  {Grenfell}, {Janson}, {Kraus}, {Krause}, {Labadie}, {Lacour}, {Line}, {Linz},
  {Loicq}, {Miguel}, {Pall{\'e}}, {Queloz}, {Rauer}, {Ribas}, {Rugheimer},
  {Selsis}, {Snellen}, {Sozzetti}, {Stapelfeldt}, {Udry}, \&
  {Wyatt}}]{quanz2022b}
{Quanz}, S.~P., {Absil}, O., {Benz}, W., {et~al.} 2022, Experimental Astronomy,
  54, 1197, \dodoi{10.1007/s10686-021-09791-z}

\bibitem[{{Rein} \& {Liu}(2012)}]{rein2012a}
{Rein}, H., \& {Liu}, S.~F. 2012, \aap, 537, A128,
  \dodoi{10.1051/0004-6361/201118085}

\bibitem[{{Rein} \& {Tamayo}(2015)}]{rein2015c}
{Rein}, H., \& {Tamayo}, D. 2015, \mnras, 452, 376,
  \dodoi{10.1093/mnras/stv1257}

\bibitem[{{Ricker} {et~al.}(2015){Ricker}, {Winn}, {Vanderspek}, {Latham},
  {Bakos}, {Bean}, {Berta-Thompson}, {Brown}, {Buchhave}, {Butler}, {Butler},
  {Chaplin}, {Charbonneau}, {Christensen-Dalsgaard}, {Clampin}, {Deming},
  {Doty}, {De Lee}, {Dressing}, {Dunham}, {Endl}, {Fressin}, {Ge}, {Henning},
  {Holman}, {Howard}, {Ida}, {Jenkins}, {Jernigan}, {Johnson}, {Kaltenegger},
  {Kawai}, {Kjeldsen}, {Laughlin}, {Levine}, {Lin}, {Lissauer}, {MacQueen},
  {Marcy}, {McCullough}, {Morton}, {Narita}, {Paegert}, {Palle}, {Pepe},
  {Pepper}, {Quirrenbach}, {Rinehart}, {Sasselov}, {Sato}, {Seager},
  {Sozzetti}, {Stassun}, {Sullivan}, {Szentgyorgyi}, {Torres}, {Udry}, \&
  {Villasenor}}]{ricker2015}
{Ricker}, G.~R., {Winn}, J.~N., {Vanderspek}, R., {et~al.} 2015, Journal of
  Astronomical Telescopes, Instruments, and Systems, 1, 014003,
  \dodoi{10.1117/1.JATIS.1.1.014003}

\bibitem[{{Roberge} {et~al.}(2017){Roberge}, {Rizzo}, {Lincowski}, {Arney},
  {Stark}, {Robinson}, {Snyder}, {Pueyo}, {Zimmerman}, {Jansen}, {Nesvold},
  {Meadows}, \& {Turnbull}}]{roberge2017}
{Roberge}, A., {Rizzo}, M.~J., {Lincowski}, A.~P., {et~al.} 2017, \pasp, 129,
  124401, \dodoi{10.1088/1538-3873/aa8fc4}

\bibitem[{{Robertson} \& {Mahadevan}(2014)}]{robertson2014b}
{Robertson}, P., \& {Mahadevan}, S. 2014, \apjl, 793, L24,
  \dodoi{10.1088/2041-8205/793/2/L24}

\bibitem[{{Robertson} {et~al.}(2014){Robertson}, {Mahadevan}, {Endl}, \&
  {Roy}}]{robertson2014a}
{Robertson}, P., {Mahadevan}, S., {Endl}, M., \& {Roy}, A. 2014, Science, 345,
  440, \dodoi{10.1126/science.1253253}

\bibitem[{{Robertson} {et~al.}(2015){Robertson}, {Roy}, \&
  {Mahadevan}}]{robertson2015b}
{Robertson}, P., {Roy}, A., \& {Mahadevan}, S. 2015, \apjl, 805, L22,
  \dodoi{10.1088/2041-8205/805/2/L22}

\bibitem[{{Robertson} {et~al.}(2020){Robertson}, {Stefansson}, {Mahadevan},
  {Endl}, {Cochran}, {Beard}, {Bender}, {Diddams}, {Duong}, {Ford}, {Fredrick},
  {Halverson}, {Hearty}, {Holcomb}, {Juan}, {Kanodia}, {Lubin}, {Metcalf},
  {Monson}, {Ninan}, {Palafoutas}, {Ramsey}, {Roy}, {Schwab}, {Terrien}, \&
  {Wright}}]{robertson2020}
{Robertson}, P., {Stefansson}, G., {Mahadevan}, S., {et~al.} 2020, \apj, 897,
  125, \dodoi{10.3847/1538-4357/ab989f}

\bibitem[{{Rosenthal} {et~al.}(2021){Rosenthal}, {Fulton}, {Hirsch},
  {Isaacson}, {Howard}, {Dedrick}, {Sherstyuk}, {Blunt}, {Petigura}, {Knutson},
  {Behmard}, {Chontos}, {Crepp}, {Crossfield}, {Dalba}, {Fischer}, {Henry},
  {Kane}, {Kosiarek}, {Marcy}, {Rubenzahl}, {Weiss}, \&
  {Wright}}]{rosenthal2021}
{Rosenthal}, L.~J., {Fulton}, B.~J., {Hirsch}, L.~A., {et~al.} 2021, \apjs,
  255, 8, \dodoi{10.3847/1538-4365/abe23c}

\bibitem[{{Scargle}(1982)}]{scargle1982}
{Scargle}, J.~D. 1982, \apj, 263, 835, \dodoi{10.1086/160554}

\bibitem[{{Schlafly} \& {Finkbeiner}(2011)}]{schlafly2011}
{Schlafly}, E.~F., \& {Finkbeiner}, D.~P. 2011, \apj, 737, 103,
  \dodoi{10.1088/0004-637X/737/2/103}

\bibitem[{{Scott} {et~al.}(2018){Scott}, {Howell}, {Horch}, \&
  {Everett}}]{scott2018}
{Scott}, N.~J., {Howell}, S.~B., {Horch}, E.~P., \& {Everett}, M.~E. 2018,
  \pasp, 130, 054502, \dodoi{10.1088/1538-3873/aab484}

\bibitem[{{Scott} {et~al.}(2021){Scott}, {Howell}, {Gnilka}, {Stephens},
  {Salinas}, {Matson}, {Furlan}, {Horch}, {Everett}, {Ciardi}, {Mills}, \&
  {Quigley}}]{scott2021}
{Scott}, N.~J., {Howell}, S.~B., {Gnilka}, C.~L., {et~al.} 2021, Frontiers in
  Astronomy and Space Sciences, 8, 138, \dodoi{10.3389/fspas.2021.716560}

\bibitem[{{Simpson} {et~al.}(2022){Simpson}, {Fetherolf}, {Kane}, {Li},
  {Pepper}, \& {Mo{\v{c}}nik}}]{simpson2022}
{Simpson}, E.~R., {Fetherolf}, T., {Kane}, S.~R., {et~al.} 2022, \aj, 163, 215,
  \dodoi{10.3847/1538-3881/ac5d41}

\bibitem[{{Skrutskie} {et~al.}(2006){Skrutskie}, {Cutri}, {Stiening},
  {Weinberg}, {Schneider}, {Carpenter}, {Beichman}, {Capps}, {Chester},
  {Elias}, {Huchra}, {Liebert}, {Lonsdale}, {Monet}, {Price}, {Seitzer},
  {Jarrett}, {Kirkpatrick}, {Gizis}, {Howard}, {Evans}, {Fowler}, {Fullmer},
  {Hurt}, {Light}, {Kopan}, {Marsh}, {McCallon}, {Tam}, {Van Dyk}, \&
  {Wheelock}}]{skrutskie2006}
{Skrutskie}, M.~F., {Cutri}, R.~M., {Stiening}, R., {et~al.} 2006, \aj, 131,
  1163, \dodoi{10.1086/498708}

\bibitem[{{Stark} {et~al.}(2020){Stark}, {Dressing}, {Dulz}, {Lopez}, {Marley},
  {Plavchan}, \& {Sahlmann}}]{stark2020}
{Stark}, C.~C., {Dressing}, C., {Dulz}, S., {et~al.} 2020, \aj, 159, 286,
  \dodoi{10.3847/1538-3881/ab8f26}

\bibitem[{{Takeda} {et~al.}(2007){Takeda}, {Ford}, {Sills}, {Rasio}, {Fischer},
  \& {Valenti}}]{takeda2007}
{Takeda}, G., {Ford}, E.~B., {Sills}, A., {et~al.} 2007, \apjs, 168, 297,
  \dodoi{10.1086/509763}

\bibitem[{{Tinney} {et~al.}(2001){Tinney}, {Butler}, {Marcy}, {Jones}, {Penny},
  {Vogt}, {Apps}, \& {Henry}}]{tinney2001}
{Tinney}, C.~G., {Butler}, R.~P., {Marcy}, G.~W., {et~al.} 2001, \apj, 551,
  507, \dodoi{10.1086/320097}

\bibitem[{{Tremaine} \& {Dong}(2012)}]{tremaine2012}
{Tremaine}, S., \& {Dong}, S. 2012, \aj, 143, 94,
  \dodoi{10.1088/0004-6256/143/4/94}

\bibitem[{{Trifonov} {et~al.}(2020){Trifonov}, {Tal-Or}, {Zechmeister},
  {Kaminski}, {Zucker}, \& {Mazeh}}]{trifonov2020}
{Trifonov}, T., {Tal-Or}, L., {Zechmeister}, M., {et~al.} 2020, \aap, 636, A74,
  \dodoi{10.1051/0004-6361/201936686}

\bibitem[{{van Leeuwen}(2007)}]{vanleeuwen2007}
{van Leeuwen}, F. 2007, \aap, 474, 653, \dodoi{10.1051/0004-6361:20078357}

\bibitem[{{Vaughan} {et~al.}(1978){Vaughan}, {Preston}, \&
  {Wilson}}]{vaughan1978}
{Vaughan}, A.~H., {Preston}, G.~W., \& {Wilson}, O.~C. 1978, \pasp, 90, 267,
  \dodoi{10.1086/130324}

\bibitem[{{Winn} \& {Fabrycky}(2015)}]{winn2015}
{Winn}, J.~N., \& {Fabrycky}, D.~C. 2015, \araa, 53, 409,
  \dodoi{10.1146/annurev-astro-082214-122246}

\bibitem[{{Wittenmyer} {et~al.}(2017){Wittenmyer}, {Jones}, {Zhao}, {Marshall},
  {Butler}, {Tinney}, {Wang}, \& {Johnson}}]{wittenmyer2017a}
{Wittenmyer}, R.~A., {Jones}, M.~I., {Zhao}, J., {et~al.} 2017, \aj, 153, 51,
  \dodoi{10.3847/1538-3881/153/2/51}

\bibitem[{{Wittenmyer} {et~al.}(2011){Wittenmyer}, {Tinney}, {O'Toole},
  {Jones}, {Butler}, {Carter}, \& {Bailey}}]{wittenmyer2011a}
{Wittenmyer}, R.~A., {Tinney}, C.~G., {O'Toole}, S.~J., {et~al.} 2011, \apj,
  727, 102, \dodoi{10.1088/0004-637X/727/2/102}

\bibitem[{{Wittenmyer} {et~al.}(2016){Wittenmyer}, {Butler}, {Tinney},
  {Horner}, {Carter}, {Wright}, {Jones}, {Bailey}, \&
  {O'Toole}}]{wittenmyer2016c}
{Wittenmyer}, R.~A., {Butler}, R.~P., {Tinney}, C.~G., {et~al.} 2016, \apj,
  819, 28, \dodoi{10.3847/0004-637X/819/1/28}

\bibitem[{{Wittenmyer} {et~al.}(2020{\natexlab{a}}){Wittenmyer}, {Wang},
  {Horner}, {Butler}, {Tinney}, {Carter}, {Wright}, {Jones}, {Bailey},
  {O'Toole}, \& {Johns}}]{wittenmyer2020b}
{Wittenmyer}, R.~A., {Wang}, S., {Horner}, J., {et~al.} 2020{\natexlab{a}},
  \mnras, 492, 377, \dodoi{10.1093/mnras/stz3436}

\bibitem[{{Wittenmyer} {et~al.}(2020{\natexlab{b}}){Wittenmyer}, {Butler},
  {Horner}, {Clark}, {Tinney}, {Carter}, {Wang}, {Johnson}, \&
  {Collins}}]{wittenmyer2020a}
{Wittenmyer}, R.~A., {Butler}, R.~P., {Horner}, J., {et~al.}
  2020{\natexlab{b}}, \mnras, 491, 5248, \dodoi{10.1093/mnras/stz3378}

\bibitem[{{Wright} {et~al.}(2010){Wright}, {Eisenhardt}, {Mainzer}, {Ressler},
  {Cutri}, {Jarrett}, {Kirkpatrick}, {Padgett}, {McMillan}, {Skrutskie},
  {Stanford}, {Cohen}, {Walker}, {Mather}, {Leisawitz}, {Gautier}, {McLean},
  {Benford}, {Lonsdale}, {Blain}, {Mendez}, {Irace}, {Duval}, {Liu}, {Royer},
  {Heinrichsen}, {Howard}, {Shannon}, {Kendall}, {Walsh}, {Larsen}, {Cardon},
  {Schick}, {Schwalm}, {Abid}, {Fabinsky}, {Naes}, \& {Tsai}}]{wright2010}
{Wright}, E.~L., {Eisenhardt}, P. R.~M., {Mainzer}, A.~K., {et~al.} 2010, \aj,
  140, 1868, \dodoi{10.1088/0004-6256/140/6/1868}

\bibitem[{{Yee} {et~al.}(2017){Yee}, {Petigura}, \& {von Braun}}]{yee2017}
{Yee}, S.~W., {Petigura}, E.~A., \& {von Braun}, K. 2017, \apj, 836, 77,
  \dodoi{10.3847/1538-4357/836/1/77}

\bibitem[{{Zechmeister} \& {K{\"u}rster}(2009)}]{zechmeister2009}
{Zechmeister}, M., \& {K{\"u}rster}, M. 2009, \aap, 496, 577,
  \dodoi{10.1051/0004-6361:200811296}

\bibitem[{{Zechmeister} {et~al.}(2018){Zechmeister}, {Reiners}, {Amado},
  {Azzaro}, {Bauer}, {B{\'e}jar}, {Caballero}, {Guenther}, {Hagen}, {Jeffers},
  {Kaminski}, {K{\"u}rster}, {Launhardt}, {Montes}, {Morales}, {Quirrenbach},
  {Reffert}, {Ribas}, {Seifert}, {Tal-Or}, \& {Wolthoff}}]{zechmeister2018}
{Zechmeister}, M., {Reiners}, A., {Amado}, P.~J., {et~al.} 2018, \aap, 609,
  A12, \dodoi{10.1051/0004-6361/201731483}

\end{thebibliography}


\end{document}